\documentclass{aa} 
\usepackage{natbib}
\usepackage{graphicx}
\usepackage{txfonts}
\usepackage{enumitem}

\begin{document} 
\title{The far-infrared/radio correlation and radio spectral index of galaxies in the
  SFR$-M_{\ast}$ plane up to
  $z$$\,\thicksim\,$$2$\thanks{\textit{Herschel} is an ESA space
    observatory with science instruments provided by European-led
    Principal Investigator consortia and with important participation
    from NASA.}}
\author{B.~Magnelli\inst{1}
	\and
	R.~J. Ivison\inst{2,3}
	\and
	D.~Lutz\inst{4}  
	\and
	I.~Valtchanov\inst{5}	    
	\and
	D.~Farrah\inst{6}
	\and
      	S.~Berta\inst{4}
	\and
	F.~Bertoldi\inst{1}
	\and
	J.~Bock\inst{7,8}
	\and
	A.~Cooray\inst{9}
	\and
	E.~Ibar\inst{10}
	\and
	A.~Karim\inst{1}
	\and
	E.~Le~Floc'h\inst{11}
	\and
	R.~Nordon\inst{12}
	\and
	S.~J.~Oliver\inst{13}
	\and
	M.~Page\inst{14}
	\and
	P.~Popesso\inst{4}
	\and
	F.~Pozzi\inst{15}
	\and
	D.~Rigopoulou\inst{16,17}
	\and
	L.~Riguccini\inst{18}
	\and
	G.~Rodighiero\inst{19}
	\and 
	D.~Rosario\inst{4}
	\and
	I.~Roseboom\inst{2}
	\and
	L.~Wang\inst{13}
	\and
	S.~Wuyts\inst{4}
        }
\institute{
Argelander-Institut
  f\"ur Astronomie, Universit\"at Bonn, Auf dem
  H\"ugel 71, D-53121 Bonn, Germany\\
  \email{magnelli@astro.uni-bonn.de} 
  \and
  Institute for Astronomy, University of Edinburgh, Blackford Hill, Edinburgh EH9 3HJ, UK
  \and
  European Southern Observatory, Karl-Schwarzschild-Str. 2, 85748 Garching bei M\"unchen, Germany
 \and
  Max-Planck-Institut f\"{u}r extraterrestrische Physik, Postfach 1312, Giessenbachstra\ss e 1, 85741 Garching, Germany
  \and
  Herschel Science Centre, ESAC, Villanueva de la Ca\~nada, 28691 Madrid, Spain 
  \and
  Department of Physics, Virginia Tech, Blacksburg, VA 24061, USA
  \and
  California Institute of Technology, 1200 E. California Blvd., Pasadena, CA 91125, USA
  \and
  Jet Propulsion Laboratory, 4800 Oak Grove Drive, Pasadena, CA 91109, USA
  \and
  Center for Cosmology, Department of Physics and Astronomy, University of California, Irvine, CA 92697, USA
  \and
  Instituto de F\'isica y Astronom\'ia, Universidad de Valpara\'iso, Avda. Gran Breta\~na 1111, Valpara\'iso, Chile
  \and
  Laboratoire AIM, CEA/DSM-CNRS-Universit{\'e} Paris Diderot, IRFU/Service
  d'Astrophysique,
  B\^at.709, CEA-Saclay, 91191 Gif-sur-Yvette Cedex, France.
  \and
  School of Physics and Astronomy, The Raymond and Beverly Sackler Faculty of Exact Sciences, Tel Aviv University, Tel Aviv 69978, Israel
  \and
  Astronomy Centre, Dept. of Physics \& Astronomy, University of Sussex, Brighton BN1 9QH, UK 
  \and
  Mullard Space Science Laboratory, University College London, Holmbury St Mary, Dorking, Surrey RH5 6NT, UK
  \and
  Dipartimento di Astronomia, Universit\`a di Bologna, via Ranzani 1, 40127 Bologna, Italy  
  \and
  Department of Physics, University of Oxford, Keble Road, Oxford, OX1 3RH, UK
  \and
  RAL Space, Science \& Technology Facilities Council, Rutherford Appleton Laboratory, Didcot, OX11 0QX, UK
  \and
  NASA Ames, Moffett Field, CA 94035, USA
  \and
  Dipartimento di Astronomia, Universit\`a di Padova, Vicolo dell'Osservatorio 3, I-35122, Italy
}

\date{Received ??; accepted ??}

\abstract{We study the evolution of the radio spectral index and
  far-infrared/radio correlation (FRC) across the star-formation rate$\,$--$\,$stellar masse (i.e.\ SFR$-M_{\ast}$) plane
  up to $z\sim 2$.  We start from a stellar-mass-selected sample of galaxies
  with reliable SFR and redshift estimates.  We
  then grid the SFR$-M_{\ast}$ plane in several redshift ranges and
  measure the infrared luminosity, radio luminosity, radio spectral
  index, and ultimately the FRC index (i.e.\ $q_{\rm FIR}$) of each
  SFR--$M_{\ast}$--$\,z$ bin.  The infrared luminosities of our
  SFR--$M_{\ast}$--$\,z$ bins are estimated using their stacked
  far-infrared flux densities inferred from observations obtained with
  the \textit{Herschel Space Observatory}.  Their radio luminosities
  and radio spectral indices (i.e.\ $\alpha$, where $S_{\nu}\propto
  \nu^{-\alpha}$) are estimated using their stacked 1.4-GHz and
  610-MHz flux densities from the Very Large Array and Giant
  Metre-wave Radio Telescope, respectively.  Our far-infrared and
  radio observations include the most widely studied blank
  extragalactic fields -- GOODS-N, GOODS-S, ECDFS, and COSMOS --
  covering a total sky area of $\sim 2.0$\,deg$^2$.  Using this
  methodology, we constrain the radio spectral index and FRC index of
  star-forming galaxies with $M_{\ast}>10^{10}$\,M$_{\odot}$ and
  $0<z<2.3$.  We find that $\alpha^{\rm 1.4\,GHz}_{\rm 610\,MHz}$ does
  not evolve significantly with redshift or with the distance of a
  galaxy with respect to the main sequence (MS) of the SFR$-M_{\ast}$
  plane (i.e.\ ${\rm \Delta log(SSFR)_{\rm
      MS}=log[SSFR(galaxy)/SSFR_{\rm MS}}(M_{\ast},z)]$).  Instead,
  star-forming galaxies have a radio spectral index consistent with a
  canonical value of 0.8, which suggests that their radio spectra are
  dominated by non-thermal optically thin synchrotron emission.  We find that the FRC
  index, $q_{\rm FIR}$, displays a moderate but statistically significant redshift evolution as $q_{\rm
    FIR}(z)$$\,=\,$$(2.35\pm0.08)\times(1+z)^{-0.12\pm0.04}$,
  consistent with some previous literature.
  Finally, we find no significant correlation between $q_{\rm FIR}$ and $\Delta$log$({\rm SSFR})_{\rm MS}$, though a weak positive trend, as observed in one of our redshift bins (i.e.\ $\Delta[q_{\rm FIR}]$/$\Delta$[$\Delta$log$({\rm SSFR})_{\rm MS}]=0.22\pm0.07$ at $0.5$$\,<\,$$z$$\,<\,$$0.8$), cannot be firmly ruled out using our dataset.
  }
\authorrunning{Magnelli et al.} \titlerunning{Far-infrared and radio
  properties of galaxies in the SFR$-M_{\ast}$ plane}
\maketitle

\section{Introduction}

The far-infrared (FIR) and radio luminosities of star-forming galaxies
are tightly related via an empirical relation, the FIR/radio
correlation \citep[FRC; e.g.][]{dejong_1985, helou_1985, helou_1988,
  condon_1992, yun_2001}.  In the local Universe, this FRC is roughly
linear across three orders of magnitude in luminosity,
$10^{9}$$\,\lesssim\,$$L_{\rm IR}(8-1000\mu{\rm m})\,[{\rm
  L_{\odot}}]$$\,\lesssim\,$$10^{12.5}$, from dwarf galaxies to
ultra-luminous infrared galaxies (ULIRGs; $L_{\rm
  IR}>10^{12}\,$L$_{\odot}$).  At lower luminosities, the FRC exhibits
some signs of non-linearity \citep[e.g.][]{bell_2003}. The FRC not only exists on galactic scales but also
holds in star-forming regions within the galaxies, down to at least 0.5\,kpc, albeit with some
variations, as shown by resolved studies of nearby galaxies
\citep[e.g.][]{beck_1988, bicay_1990, murphy_2008, dumas_2011,
  tabatabaei_2013b, tabatabaei_2013a}.

The FRC is thought to be caused by star-formation activity in
galaxies. Young massive stars ($\gtrsim\,$$8\,$M$_{\odot}$) radiate
most of their energy at UV wavelengths, photons that dominate the UV continuum luminosity of galaxies and which are absorbed and
re-emitted in the FIR regime by dust.  After several Myrs,
these young massive stars explode as supernov\ae\ (SNe), accelerating
cosmic rays (CRs) into the general magnetic field of their host
galaxy and resulting in diffuse synchrotron emission.  Averaged over a
star-formation episode, massive stars thus provide a common origin for
the FIR and synchrotron emission of galaxies.  This shared
origin is the essence of all models constructed to explain the FRC on
both global scales (e.g.\ the `calorimeter model', V\"olk
\citeyear{volk_1989}; the `conspiracy model', Lacki et al.
\citeyear{lacki_2010a}) and local scales (e.g.\ `small-scale
dynamo', Schleicher \& Beck \citeyear{schleicher_2013}; see also
Niklas \& Beck \citeyear{niklas_1997}). While many of these models are
able to reproduce the basic properties of the FRC, none are consistent
with all constraints provided by observations.

Despite our lack of a thorough theoretical understanding, the very
tight empirical FRC (with a dispersion of $\thicksim\,$$0.25\,$dex; e.g.\ Yun et al. \citeyear{yun_2001}) has been used extensively for a number of
astrophysical purposes.  For example, the FRC observed in the local Universe has been exploited to empirically
calibrate the radio luminosity as a star-formation rate (SFR)
indicator, using the known $L_{\rm IR}\,$--$\,$SFR relation for
galaxies whose activities are not dominated by active galaxy nuclei
\citep[AGN; e.g.][]{condon_1992,bell_2003,murphy_2011b}.  This allows
us to estimate the level of star formation in high-redshift galaxies,
taking advantage of radio observations that are often deeper than
FIR surveys with better spatial resolution.  The locally observed FRC has
also been used to identify samples of radio-loud AGNs and to study
their properties \citep[e.g.][]{donley_2005,park_2008,delmoro_2013}.
Finally, at high redshift, the local FRC has been used to estimate the
distance \citep[e.g.][]{carilli_1999} or the dust temperature
\citep[e.g.][]{chapman_2005} of luminous starbursts, namely the
submillimetre galaxies \citep[SMGs --][]{smail_1997}. As these
examples demonstrate, the use of the FRC has become an important tool for
extragalactic astrophysics. Upcoming surveys with the Jansky Very Large Array (JVLA) will make its
applications even more important.

Although the FRC is characterised well at low redshift, its form and
thus its applicability at high redshift still have to be firmly demonstrated.
From a theoretical point of view, 
we expect the FRC to break down at high redshift (i.e.\ $z$$\,\gtrsim\,$2$\,$--$\,$3) because CR electron
cooling via inverse Compton (IC) scattering off the cosmic microwave
background (CMB; $U_{\rm CMB}$$\,\propto\,$$(1+z)^{4}$) photons is supposed to dominate over synchrotron
cooling \citep[e.g.][]{murphy_2009b,lacki_2010b,schleicher_2013}.
However, the exact redshift and amplitude of this breakdown varies
with models and with the assumed properties of high-redshift galaxies.
From the observational point of view, the characteristics of the FRC at
high redshift have been subject to extensive debate.
Some studies have found that the FRC stays unchanged or suffers only minor
variations at high redshift
\citep[e.g.][]{appleton_2004,ibar_2008,bourne_2011} and others
have found significant evolution of the FRC in the bulk of the
star-forming galaxy population \citep[e.g.][]{seymour_2009} or
for a subsample of it \citep[e.g.\ SMGs;][]{murphy_2009}. 
\citet{sargent_2010} argue
that these discrepant measurements could be explained by
selection biases due to the improper treatment of flux limits from
non-detections amongst radio- and FIR-selected samples. Then, applying a survival analysis to properly treat these non-detections, they conclude that the
FRC remains unchanged or with little variations out to
$z$$\,\thicksim\,$$1.4$.  However, the results of  \citet{sargent_2010} were still limited by relatively sparse coverage of the
FIR and radio spectral energy distributions (SEDs) of high-redshift galaxies. 
Noticing these limitations, \citet{ivison_2010a}
studied the evolution of the FRC using FIR (250, 350 and 500$\,\mu$m)
observations from BLAST \citep{devlin_2009} and multi-frequency radio
observations (1.4\,GHz and 610\,MHz).  Starting from a
mid-infrared-selected sample and accounting for radio non-detections
using a stacking analysis, they found an evolution of the FRC with
redshift as $\propto(1+z)^{-0.15\pm0.03}$. Repeating a similar
analysis using early observations from the \textit{Herschel Space
  Observatory}, \citet{ivison_2010b} found support for such redshift
evolution in the FRC.
These findings demand modifications to any high-redshift results that adopted the local FRC.
However, because results from \citet{ivison_2010a,ivison_2010b} were based on mid-infrared-selected samples, they might
still be affected by some selection biases.

In this paper, we aim to study the FRC across $0<z<2.3$, avoiding the
biases mentioned above. To obtain a good FIR and radio spectral coverage,  we use deep FIR (100, 160, 250, 350 and 500$\,\mu$m) observations from the \textit{Herschel Space Observatory} \citep{pilbratt_2010} and deep radio 1.4-GHz VLA and 610-MHz Giant Metre-wave Radio Telescope (GMRT) observations.
To minimise selection biases and properly account for radio and FIR non-detections, we use a careful stacking analysis based on the positions of stellar-mass-selected samples of galaxies.
These stellar-mass-selected samples are complete for star-forming galaxies down to $10^{10}\,$M$_{\odot}$ across $0<z<2.3$ and are built from the large wealth of multi-wavelength observations available for the blank extragalactic fields used in our analysis -- GOODS-N, GOODS-S, ECDFS and COSMOS.

Besides investigating the evolution of the FRC as a function of redshift, we also aim to study its evolution in the SFR--stellar mass ($M_{\ast}$) plane.
Indeed, recent results have shown that the physical properties (e.g.\ morphology, CO-to-H$_{2}$ conversion factor, dust
temperature) of star-forming galaxies vary with their positions in the SFR-$M_{\ast}$ plane (e.g.\ Wuyts et al.\ \citeyear{wuyts_2011b}, Magnelli et al.\ \citeyear{magnelli_2012b}, Magnelli et al.\ \citeyear{magnelli_2014}).
In particular, these properties correlate with the distance of a galaxy to the so-called `main sequence' (MS) of the SFR-$M_{\ast}$ plane, i.e.\ the sequence where the bulk of the star-forming galaxy population resides and which is characterised by SFR$\,\propto$$\,M_{\ast}^\gamma$, with $0.5<\gamma<1.0$ \citep[][]{brinchmann_2004, schiminovich_2007,noeske_2007a,elbaz_2007, daddi_2007a, pannella_2009, dunne_2009,rodighiero_2010b, oliver_2010, karim_2011, mancini_2011, whitaker_2012}.
Studying the FRC as a function of the distance of a galaxy from the MS (i.e.\ ${\rm \Delta log(SSFR)_{\rm MS}=log[SSFR(galaxy)/SSFR_{\rm MS}}(M_{\ast},z)]$) will allow us to estimate if this correlation evolves from normal star-forming galaxies (${\rm \Delta log(SSFR)_{\rm MS}}$$\thicksim$$0$) to starbursting galaxies (${\rm \Delta log(SSFR)_{\rm MS}}$$\thicksim$$1$), as suggested by some local observations (e.g.\ Condon et al. \citeyear{condon_1991}; but see, e.g.\ Yun et al. \citeyear{yun_2001}).
\begin{table*}
\scriptsize
\caption{\label{tab:data} Main properties of the PEP/GOODS-H/HerMES observations used in this study.}
\begin{center}
\begin{tabular}{ c c c  ccc  c cccc} 
\hline \hline
& & &\multicolumn{3}{c}{\rule{0pt}{3ex} PACS} && \multicolumn{4}{c}{SPIRE} \\
\cline{4-6} \cline{8-11}
\multicolumn{2}{c}{\rule{0pt}{2.5ex}Field}  && Eff.\ area  & 100-$\mu$m 3$\sigma^\mathrm{\,a}$ &  160-$\mu$m 3$\sigma^\mathrm{\,a}$ && Eff.\ area &  250-$\mu$m $3\sigma^\mathrm{\,a}$ & 350-$\mu$m $3\sigma^\mathrm{\,a}$ & 500-$\mu$m $3\sigma^\mathrm{\,a}$ \\
 & & & arcmin$^2$  & mJy  & mJy  && arcmin$^2$  & mJy  & mJy  & mJy \\
\hline
\rule{0pt}{3ex}GOODS-S$^{\rm b}$ & ${\rm 03^{h}32^{m},\,-27^{\circ}48\arcmin}$  && 200 (100) &  1.0 (0.6)  &  2.1 (1.3) &&  400 &  7.8 &  9.5 &  12.1 \\ 
ECDFS & ${\rm 03^{h}32^{m},\,-27^{\circ}48\arcmin}$  && 900 &  4.4 &  8.3 &&  900 &  7.8 &  9.1 &  11.9 \\ 
GOODS-N & ${\rm 12^{h}36^{m},\,+62^{\circ}14\arcmin}$ && 200 &  1.0 &  2.1 &&  900 & 9.2 & 12 & 12.1 \\ 
COSMOS & ${\rm 00^{h}00^{m},\,+02^{\circ}12\arcmin}$  && 7344 & 5.0 &  10.2 && 7225 &  8.1 &  10.7 &  15.4 \\ 
\hline
\end{tabular}
\end{center}
\textbf{Notes:}  $\ ^{\mathrm{a}}$ r.m.s.\ values include confusion noise. $^{\rm b}$ Values in parentheses correspond to the PACS `ultradeep' part of the GOODS-S field, as obtained by combining GOODS-H and PEP observations \citep{magnelli_2013}.
\end{table*}

To study the evolution of the FRC as a function of redshift and as a function of the distance of a galaxy from the MS (i.e.\ ${\rm \Delta log(SSFR)_{\rm
    MS}}$), we grid the
SFR$-M_{\ast}$ plane in several redshift ranges and estimate for each
SFR--$M_{\ast}$--$\,z$ bin its infrared luminosity, radio spectral
index (i.e.\ $\alpha$, where $S_{\nu}\propto \nu^{-\alpha}$) and radio luminosity using a FIR and radio stacking
analysis.  Thanks to this methodology, we are able to statistically and accurately
constrain the radio spectral index and FRC of all star-forming
galaxies with $M_{\ast}>10^{10}\,$M$_{\odot}$, $\Delta$log$({\rm
  SSFR})_{\rm MS}$$\,>\,$$-0.3$ and $0<z<2.3$.
  
The paper is structured as follows.  In Section \ref{sec:data} we
present the \textit{Herschel}, VLA and GMRT observations used in this
study, as well as our stellar-mass-selected sample.  Section \ref{sec:dust
  analysis} presents the FIR and radio stacking analysis used to infer
the infrared luminosity, radio spectral index and radio luminosity of
each of our SFR--$M_{\ast}$--$\,z$ bins.  Evolution of the radio
spectral index, $\alpha^{\rm 1.4\,GHz}_{\rm 610\,MHz}$, and of the FRC
with ${\rm \Delta log(SSFR)_{\rm MS}}$ and redshift are presented in
Section \ref{subsec:spectral index} and \ref{subsec:q}, respectively.
Results are discussed in Section \ref{sec:discussion} and summarised
in Section \ref{sec:conclusion}.

Throughout the paper we use a cosmology with
$H_{0}$$\,=\,$$71\,\rm{km\,s^{-1}\,Mpc^{-1}}$,
$\Omega_{\Lambda}$$\,=\,$$0.73$ and $\Omega_{\rm M}$$\,=\,$$0.27$.

\section{Data\label{sec:data}}

\subsection{Herschel observations}

To study the FIR properties of galaxies in the SFR$-M_{\ast}$ plane,
we used deep FIR observations of the COSMOS, GOODS-N, GOODS-S and
ECDFS fields provided by the \textit{Herschel Space observatory}.
Observations at 100 and 160$\,\mu$m were obtained by the Photodetector
Array Camera and Spectrometer \citep[PACS;][]{poglitsch_2010} as part
of the PACS Evolutionary Probe
\citep[PEP\footnote{\texttt{http://www.mpe.mpg.de/ir/Research/PEP}};][]{lutz_2011}
guaranteed time key programme and the GOODS-\textit{Herschel}
\citep[GOODS-H\footnote{\texttt{http://hedam.oamp.fr/GOODS-Herschel}};][]{elbaz_2011}
open time key programme.  Observations at 250, 350 and 500$\,\mu$m
were obtained by the Spectral and Photometric Imaging Receiver
\citep[SPIRE][]{griffin_2010} as part of the \textit{Herschel}
Multi-tiered Extragalactic Survey
\citep[HerMES\footnote{\texttt{http://hermes.sussex.ac.uk}}][]{oliver_2012}.
The PEP, PEP/GOODS-H and HerMES surveys and data reduction methods are
described in \citet{lutz_2011}, \citet[][see also Elbaz et al.\
\citeyear{elbaz_2011}]{magnelli_2013} and \citet{oliver_2012},
respectively.  Here we only summarise the information relevant for our
study.

\textit{Herschel} flux densities were derived using a
point-spread-function-fitting method, guided by the known position of
sources detected in deep 24-$\mu$m observations from the Multiband
Imaging Photometer \citep[MIPS;][]{rieke_2004} on board the
\textit{Spitzer Space Observatory}.  This method provides reliable and
highly complete PACS/SPIRE source catalogues \citep{lutz_2011,
  magnelli_2013}.  We note that the small fraction of
\textit{Herschel} sources without a MIPS counterpart and thus missing in our source catalogues
\citep{magdis_2011} will be considered via our stacking analysis which is based on the positions of complete stellar-mass-selected samples (Section~\ref{subsec:FIR stacking}).  The extraction of PACS
sources was accomplished using the method described in
\citet{magnelli_2013}, while for SPIRE sources it was done using the
method described in \citet{roseboom_2010}, both using the same
24-$\mu$m catalogues.  In GOODS-S/N, we used the GOODS
24-$\mu$m catalogue, reaching a 3-$\sigma$ limit of 20$\,\mu$Jy
\citep{magnelli_2009,magnelli_2011a}; in the ECDFS, we used the FIDEL
24-$\mu$m catalogue, reaching a 3-$\sigma$ limit of 70$\,\mu$Jy
\citep{magnelli_2009}; in COSMOS, we used a 24-$\mu$m
catalogue with a 3-$\sigma$ limit of 45$\,\mu$Jy
\citep{lefloch_2009}.  Reliability, completeness and contamination of
our PACS/SPIRE catalogues were tested using Monte Carlo simulations.
Table~\ref{tab:data} summarises the depths of all these catalogues.
Note that SPIRE observations of GOODS-N/S cover much larger effective areas than those from PACS (see Tab.~\ref{tab:data}). However, here we restricted our study to regions with both PACS and SPIRE data.

Our MIPS-PACS-SPIRE catalogues were cross-matched with our multi-wavelength catalogues (Sect.~\ref{subsec:multi-sample}), using their MIPS/IRAC positions and a matching radius of 1\arcsec.

\begin{table}
\scriptsize
\caption{\label{tab:data radio} Main properties of the $1.4\,$GHz-VLA and $610\,$MHz-GMRT observations used in this study.}
\begin{center}
\begin{tabular}{ c cc  ccc } 
\hline \hline
 &\multicolumn{2}{c}{\rule{0pt}{3ex} $\,1.4\,$GHz$\,$--$\,$VLA} && \multicolumn{2}{c}{$\,610\,$MHz$\,$--$\,$GMRT} \\
\cline{2-3} \cline{5-6}
\rule{0pt}{2.5ex}Field  & Eff.\ area  & $1\sigma$ && Eff.\ area &  $1\sigma$ \\
 & arcmin$^2$  & $\mu$Jy  && arcmin$^2$  & $\mu$Jy   \\
\hline
\rule{0pt}{3ex}ECDFS \& GOODS-S & 1156 &  8 && 4900 &  15  \\ 
GOODS-N & 225 &  $3.9\,$--$\,8$ && 400 &  15  \\ 
COSMOS$^{\rm a}$ &  14400 (3600) & 15 (10) && N/A & N/A  \\ 
\hline
\end{tabular}
\end{center}
\textbf{Notes:} $^{\rm a}$ Values in parentheses correspond to the deep VLA observations of the centre of the COSMOS field, as obtained by combining data from the VLA-COSMOS Large and Deep project \citep{schinnerer_2010}.
\end{table}

\subsection{VLA observations}

To study the radio properties of our galaxies, we use deep publicly
available 1.4-GHz VLA observations.  In the COSMOS field, we use
observations from the VLA-COSMOS
Project\footnote{\texttt{http://www.mpia-hd.mpg.de/COSMOS/}}
\citep{schinnerer_2004,schinnerer_2007,schinnerer_2010}.  This project
has imaged the entire COSMOS field at 1.4\,GHz to a mean r.m.s.\ noise
of $\thicksim$10$\,$(15)$\,$$\,\mu$Jy\,beam$^{-1}$ across its central
1$\,$(2)$\,$deg$^{2}$ with a resolution of
$1\farcs5$$\,\times\,$$1\farcs4$ (FWHM of the synthesised beam).  In
the GOODS-N field, we use deep 1.4-GHz VLA
observations\footnote{\texttt{http://www.stsci.edu/science/goods/}}
described in \citet{morrison_2010}.  This map has a synthesised beam
size of $1\farcs7$$\,\times\,$$1\farcs7$ and a mean r.m.s.\ noise of
$\thicksim\,$3.9$\,\mu$Jy\,beam$^{-1}$ near its centre, rising to
$\thicksim\,$8$\,\mu$Jy\,beam$^{-1}$ at a distance of $15\arcmin$ from
the field centre.  Finally, in the GOODS-S and ECDFS fields, we use
deep 1.4-GHz VLA
observations\footnote{\texttt{http://www.astro.umd.edu/$\sim$nmiller/VLA\_ECDFS.html}}
presented in \citet{miller_2008,miller_2013}.  This map has a
synthesised beam size of $2\farcs8$$\,\times\,$$1\farcs6$ and an
mean r.m.s.\ noise of $\thicksim$8$\,\mu$Jy\,beam$^{-1}$.
Table~\ref{tab:data radio} summarises the properties of all these
observations.

\subsection{GMRT observations\label{subsec:GMRT}}

GMRT 610-MHz observations of the GOODS-N and ECDFS fields were
obtained in 2009--10 as part of the `GOODS-50' project (PI: Ivison).
The ECDFS observations were described by \citet{thomson_2014}; those
of GOODS-N covered a single pointing rather than the six used to cover
ECDFS, but in most other respects were identical.
The GMRT observations of ECDFS and GOODS-N reach a
mean r.m.s.\ noise of $\thicksim$15$\,\mu$Jy\,beam$^{-1}$.
The synthesised beam measured $4\farcs4$$\,\times\,$$3\farcs0$ and 
$5\farcs0$$\,\times\,$$4\farcs8$ in the ECDFS and GOODS-N maps, 
respectively.
Table~\ref{tab:data radio} summarises the properties of all these
observations.

Note that there are no GMRT observations of the COSMOS 
field. Therefore, constraints on the radio spectral index of star-forming galaxies
 (Sect. \ref{subsec:spectral index}) are obtained from a somehow more limited 
 galaxy sample, covering a sky area of $\thicksim\,$$0.3\,$deg$^2$.
These limitations could have restrained our ability to constrain the radio spectral index of faint star-forming galaxies.
 However, we demonstrate in Sect.~\ref{subsec:completeness} that with our current GMRT observations, we are able to estimate the radio spectral index of galaxies with $M_{\ast}>10^{10}\,$M$_{\odot}$, $\Delta$log$({\rm SSFR})_{\rm MS}$$\,>\,$$-0.3$ and $0<z<2.3$. 
 These constraints are sufficient for the purpose of our study.
In addition, because we do not expect significant cosmic variance, we can apply these constraints obtained on a sky area of $\thicksim\,$$0.3\,$deg$^2$ to our entire galaxy sample.

\subsection{Multi-wavelength catalogues\label{subsec:multi-sample}}

The large wealth of multi-wavelength data used in our study is
described in detail in \citet[][; see also Magnelli et al.\
\citeyear{magnelli_2014}]{wuyts_2011a, wuyts_2011b}.  Here we only
summarise the properties relevant for our study.

In the COSMOS field we used 36 medium and broad-band observations
covering the optical to near-infrared SEDs of galaxies
\citep{ilbert_2009, gabasch_2008}.  We restricted these catalogues to
$i$$\,<\,$$25$ and to sources not flagged as problematic in the
catalogue of \citet{ilbert_2009}.  In the GOODS-S field, we used the
$K_s$$\,<\,$$24.3$ (5$\sigma$) FIREWORKS catalogue, providing
photometry in 16 bands from $U$ to IRAC wavelengths.  In the GOODS-N
field, we used the $K_s$$\,<\,$$24.3$ (3$\sigma$) catalogue created as
part of the PEP survey\footnote{publicly available at
  \texttt{http://www.mpe.mpg.de/ir/Research/PEP}}, which provides
photometry in 16 bands from \textit{GALEX} to IRAC wavelengths.
Finally, in the ECDFS field, we used the $R$$\,<\,$$25.3$ (5$\sigma$)
catalogue described in \citet{cardamone_2010}.  This multi-wavelength
catalogue provides photometry in 18 bands from $U$ to IRAC
wavelengths.  In the following, ECDFS only corresponds to the
outskirts of the original ECDFS region, i.e.\ when we refer to the
ECDFS field, we implicitly exclude the central GOODS-S region.

\begin{figure}
\center
\includegraphics[width=9.2cm]{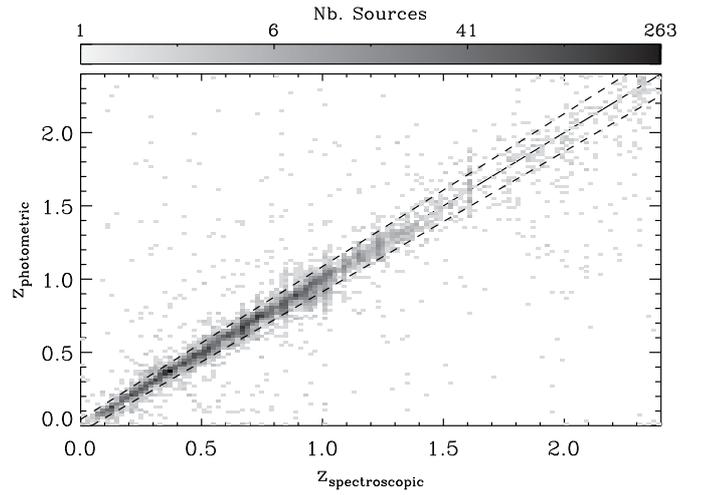}
\caption{\label{fig:zspec zphot} Comparison between the photometric and spectroscopic redshifts for the $12\,132$ galaxies with both kinds of redshifts in our multi-wavelength catalogues. Dashed lines represent three times the median absolute deviation found in the redshift range of our study (i.e.\ $\Delta z/(1+z)$$\,=\,$$0.014$ at $0$$\,<\,$$z$$\,<\,$$2.3$).}
\end{figure}

Spectroscopic redshifts for our galaxies were taken from a combination
of various studies \citep{cohen_2000, cristiani_2000, croom_2001,
  cimatti_2002, wirth_2004, cowie_2004, lefevre_2004, szokoly_2004,
  van_der_wel_2004, mignoli_2005, vanzella_2006, vanzella_2008,
  reddy_2006, barger_2008, cimatti_2008, kriek_2008, lilly_2009,
  treister_2009, balestra_2010}.  For sources without a spectroscopic
redshift, we used photometric redshift computed using \texttt{EAZY}
\citep{brammer_2008} exploiting all the available
optical/near-infrared data.  The quality of these photometric
redshifts was assessed via comparison with spectroscopically confirmed
galaxies.  The median and median absolute deviation of $\Delta z/(1+z)$ are ($-0.001$;
$0.013$) at $z$$\,<\,$$1.5$, ($-0.007$; $0.066$) at
$z$$\,>\,$$1.5$ and ($-0.001$; $0.014$) at $0$$\,<\,$$z$$\,<\,$$2.3$ (Fig. \ref{fig:zspec zphot}).

\subsubsection{Stellar masses}

The stellar masses of our galaxies were estimated by fitting all
$\lambda_{\rm obs}\leqslant8\,\mu$m data to \citet{bruzual_2003}
templates using \texttt{FAST} (Fitting and Assessment of Synthetic
Templates; Kriek et al. \citeyear{kriek_2009}).  The rest-frame
template error function of \citet{brammer_2008} was used to
down-weight data points with $\lambda_{{\rm rest}}\geqslant2\,\mu$m.
For those stellar mass estimates we adopted a \citet{chabrier_2003} IMF.
Full details on those estimates and their limitations are given in \citet{wuyts_2011a,wuyts_2011b}.
\begin{figure*}
\center
\includegraphics[width=14.9cm]{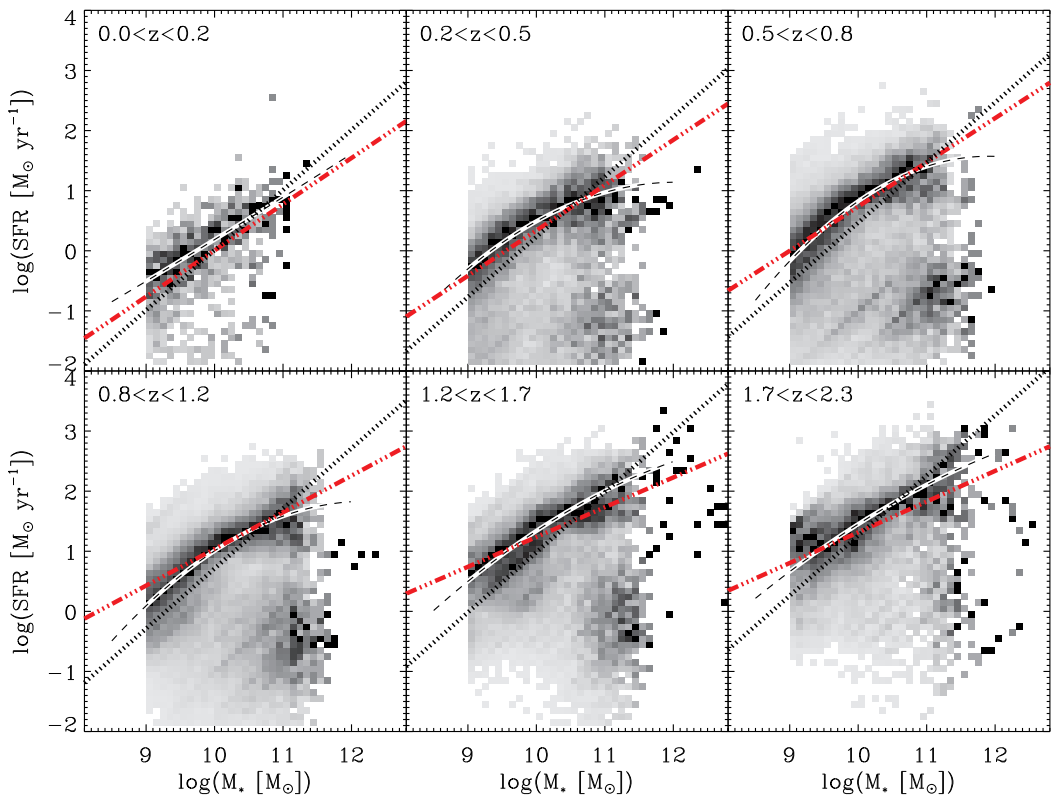}
\caption{\label{fig:sfr mstar} Number density of sources in the
  SFR$-M_{\ast}$ plane.  Shading is independent for each stellar mass
  bin, i.e.\ the darkest colour indicates the highest number density
  of sources in the stellar mass bin and not the highest number
  density of sources in the entire SFR$-M_{\ast}$ plane.  Short-dashed
  lines on a white background show the second-order polynomial
  functions used here to describe the MS of star formation \citep{magnelli_2014}.  Dotted
  lines represent the MS and its redshift evolution as found in \citet{elbaz_2011}.  The red
  triple-dot-dashed lines represent the MS and its redshift evolution as found in
  \citet{rodighiero_2010b}.  }
\end{figure*}

\subsubsection{Star-formation rates}

To estimate the SFRs of our galaxies we used the cross-calibrated
`ladder of SFR indicators' established in \citet{wuyts_2011a}.  This
uses the best indicator available for each galaxy and establishes a
consistent scale across all of them.  For galaxies only detected in
the rest-frame UV (i.e.\ those without a mid- or far-infrared
detection), SFRs were estimated from the best fits obtained with
\texttt{FAST}.  For galaxies with detections both in the rest-frame UV
and the mid-/far-infrared, SFRs were estimated by combining the
unobscured and re-emitted emission from young stars.  This was done
following \citet{kennicutt_1998} and adopting a \citet{chabrier_2003}
IMF:
\begin{equation}
\label{eq: kennicutt}
{\rm SFR_{UV+IR}[M_{\odot}\,yr^{-1}]}=1.09\times10^{-10}\,(L_{{\rm IR}}\,[{\rm L}_{\odot}]+3.3\times L_{{\rm 2800}}\,[{\rm L}_{\odot}]),
\end{equation}
where $L_{2800}$$\,\equiv\,$$\nu L_{\nu}(2800\,\AA)$ was computed with
\texttt{FAST} from the best-fitting SED and the rest-frame infrared
luminosity $L_{{\rm IR}}$$\,\equiv\,$$L[8-1000\,\mu {\rm m}]$ is
derived from the mid-/far-infrared observations.  For galaxies with
FIR detections, $L_{{\rm IR}}$ was inferred by fitting their FIR flux
densities (i.e.\ those measured using PACS and SPIRE) with the SED
template library of \citet[][DH]{dale_2002}, leaving the normalisation
of each SED template as a free parameter\footnote{using the SED
  template library of \citet{chary_2001} instead of that of DH has no
  impact on our results.}.  The infrared luminosities of galaxies with
only a mid-infrared detection were derived by scaling the SED template
of MS galaxies \citep{elbaz_2011} to their 24-$\mu$m flux
densities.  \citet{magnelli_2014} have shown that this specific SED
template provides accurate 24$\,\mu$m-to-$L_{{\rm IR}}$ conversion
factors for such galaxies.

\subsubsection{Active galactic nuclei contamination}
Active Galactic Nuclei (AGNs) can affect the observed FRC of
star-forming galaxies.  Such AGNs must thus be excluded from our
sample.  To test for the presence of AGNs, we used the deepest
available \textit{Chandra} and \textit{XMM-Newton} X-ray observations,
identifying AGNs as galaxies with $L_{\rm X}\, [0.5-8.0\ {\rm
  keV}]$$\,\ge\,$$3\times10^{42}\,\rm{erg\, s^{-1}}$ and $L_{\rm X}\,
[2.0-10.0\ {\rm keV}]$$\,\ge\,$$3\times10^{42}\,\rm{erg\, s^{-1}}$,
respectively \citep{bauer_2004}.  In the GOODS-N and -S fields, X-ray
observations were taken from the \textit{Chandra} 2-Ms catalogues of
\citet{alexander_2003} and \citet{luo_2008}, respectively.  In the
COSMOS field, we used the \textit{XMM-Newton} catalogue of
\citet{cappelluti_2009}.  Finally, for the ECDFS, we used the 250-ks
\textit{Chandra} observations, which flank the 2-Ms CDFS observations
\citep{lehmer_2005}.  All these X-ray-selected AGNs have been removed
from our sample.  Note that radio-loud AGNs without X-ray detection are excluded
in our radio stacking procedure, through use of median stacking (Section~\ref{subsec:radio stacking}).

\subsection{Final sample}

Our four multi-wavelength catalogues are not homogeneously selected
and are not uniform in depth, which naturally translates into
different completeness limits in the SFR$-M_{\ast}$ plane.  These
issues have been discussed and studied in \citet{magnelli_2014}.  They
found that in the GOODS-S, GOODS-N and COSMOS fields our
multi-wavelength catalogues are complete for star-forming galaxies with
$M_{\ast}$$\,>\,$$10^{10}\,$M$_{\odot}$ up to $z$$\,\thicksim\,$$2$.
Because the ECDFS multi-wavelength catalogue is based on deeper
optical/near-infrared observations than those of the COSMOS field, we
conclude that this catalogue also provides us with a complete sample
of star-forming galaxies with $M_{\ast}$$\,>\,$$10^{10}\,$M$_{\odot}$ up to
$z$$\,\thicksim\,$$2$.  In the rest of the paper, we restrict our
results and discussion to galaxies with
$M_{\ast}$$\,>\,$$10^{10}\,$M$_{\odot}$.

Our final sample contains 8$\,$846, 4$\,$753, 66$\,$070, and
254$\,$749 sources in the GOODS-N, GOODS-S\footnote{Although the
  GOODS-S and -N multi-wavelength catalogues both correspond to
  $K_{\rm s}$$\,<\,$$24.3$, the GOODS-S catalogue contains fewer
  sources than the GOODS-N catalogue because it includes only sources
  with $>\,$5$\sigma$ while the GOODS-N multi-wavelength catalogue
  extends down to a significance of 3$\sigma$.}, ECDFS and COSMOS
fields, respectively.  Of these sources, 29\%, 26\%, 1\% and 3\% have
a spectroscopic redshift, while the rest have photometric redshift
estimates.  Because we are studying the FRC, the SFR--$M_{\ast}$--$\,z$ bins that enter our analysis (see Sect.~\ref{sec:dust analysis}) are generally dominated by sources that have individual mid-infrared (and for part of them far-infrared) detections.  In
GOODS-N, GOODS-S, ECDFS and COSMOS, 19\%, 28\%, 9\% and 12\% of the
galaxies have mid- or far-infrared detections, respectively.  Among
those sources, 60\%, 45\%, 5\% and 11\% have a spectroscopic redshift.

The Fig.~\ref{fig:sfr mstar} shows the number density of sources in the
SFR$-M_{\ast}$ plane.  Over a broad range of stellar masses,
star-forming galaxies (i.e.\ excluding massive and passive galaxies
situated in the lower right part of the SFR$-M_{\ast}$ plane) follow a
clear SFR$-M_{\ast}$ correlation.  This correlation is known as the
`MS of star formation' \citep{noeske_2007a}.  In the rest of the
paper, we parametrise this MS using second-order polynomial functions
as derived by \citet[][see their Table~2]{magnelli_2014}.  These
functions are presented in Fig.~\ref{fig:sfr mstar}.  Comparisons between this parametrisation and
those from the literature are presented and discussed in
\citet{magnelli_2014}.  Briefly, the MS observed in our sample is
consistent with the literature at
$M_{\ast}$$\,>\,$$10^{10}\,$M$_{\odot}$, i.e.\ within the stellar mass
range of interest for our study.

\section{Data analysis\label{sec:dust analysis}}

The aim of this paper is to study the evolution of the FRC and radio
spectral index with redshift and with respect to the position of
galaxies in the SFR$-M_{\ast}$ plane.  Although we could base this
analysis on galaxies individually detected at FIR and/or radio
wavelengths, such an approach would be subject to strong limitations,
mainly due to complex selection functions \citep[see
e.g.][]{sargent_2010}.  Instead, we adopted a different approach based
on a careful FIR and radio stacking analysis of a stellar-mass-selected
sample.  This allows us to probe the properties of the FRC and radio
spectral index, delving well below the detection limits of current FIR
and radio observations.
Of course, the use of this stacking analysis has the obvious drawback that one can only study the mean properties of the FRC within the SFR$-M_{\ast}$ plane, while outliers are completely missed out.
This limitation has to be taken into account while discussing our results.
In addition, our stacking analysis has to be performed with great care, especially for the FIR observations where large beam sizes might lead to significant flux biases if the stacked samples are strongly clustered.

\subsection{Determination of far-infrared properties through stacking\label{subsec:FIR stacking}}

To estimate the FIR properties (i.e.\ $L_{\rm FIR}$ and $T_{\rm
  dust}$) of a given galaxy population, we stacked their
\textit{Herschel} observations.  The stacking method adopted here is
similar to that used in \citet{magnelli_2014}.  In the following we
only summarise the key steps of this method, while for a full
description we refer the reader to \citet{magnelli_2014}.

For each galaxy population (i.e.\ for each SFR--$M_{\ast}$--$\,z$ bin)
and for each \textit{Herschel} band, we stacked the residual image
(original maps from which we removed all 3-$\sigma$ detections) at the
positions of undetected sources (i.e.\ sources with
$S_{\textit{Herschel}}$$\,<\,$$3\sigma$).  The stacked stamp of each
galaxy was weighted with the inverse of the square of the error map.
The flux densities of the final stacked images were measured by
fitting with the appropriate PSF.  Uncertainties on these flux
densities were computed by means of a bootstrap analysis.  The mean
flux density ($S_{{\rm bin}}$) of the corresponding galaxy population
was then computed by combining the fluxes of undetected and detected
sources:
\begin{center}
\begin{equation}\label{eq:mean stack} 
S_{{\rm bin}}=\frac{m\times S_{{\rm stack}}+\sum_{i=1}^{n}S_{i}}{n+m},
\end{equation}
\end{center}
where $S_{{\rm stack}}$ is the stacked flux density of the $m$
undetected sources, and $S_{i}$ is the flux density of the $i$-th
detected source (out of a total of $n$).  We note that consistent
results are obtained if we repeat the stacking analysis using the
original PACS/SPIRE maps and combining all sources in a given
SFR--$M_{\ast}$--$\,z$ bin, regardless of whether they are
individually detected or not.  We have verified that consistent
results are obtained via a median stacking method, rather than the
mean stacking described above.

To verify that the clustering properties of our stacked samples have
no significant effect on our stacked FIR flux densities, we used
simulations from \citet[][]{magnelli_2014}. Briefly, simulated
\textit{Herschel} flux densities of all sources in our final sample
were estimated using the MS template of \citet{elbaz_2011}, given
their observed redshifts and SFRs.  Simulated \textit{Herschel} maps with `real' clustering properties were then produced using the observed positions and simulated
\textit{Herschel} flux densities of each sources of our fields. Whenever we stacked a given
galaxy population on the real \textit{Herschel} images, we also
stacked at the same positions the simulated images and thus obtained a
simulated stacked flux density ($S^{\rm simu}_{\rm stack}$).  Then we
compared the $S^{\rm simu}_{\rm stack}$ with the expected mean flux
density of this simulated population, i.e.\ $S_{\rm stack}^{\rm
  expected}$.  If ABS(($S_{{\rm stack}}^{{\rm simu}}-S_{{\rm
    stack}}^{{\rm expected}}$)/$S_{{\rm stack}}^{{\rm
    expected}}$)$\,>\,$$0.5$, then the real stacked flux densities
were identified as being potentially affected by clustering.  This
$0.5$ value was empirically defined as being the threshold above which
the effect of clustering would not be captured within the flux
uncertainties of our typical S/N$\,\thicksim\,$4 stacked flux
densities.  The largest clustering effects are observed at low flux
densities and in the SPIRE 500-$\mu$m band, as expected.
More details on these simulations can be found in section 3.2.2 of \citet{magnelli_2014}.

From the mean \textit{Herschel} flux densities of each galaxy
population we inferred their rest-frame FIR luminosities (i.e.\
$L_{\rm FIR}$, where $L_{\rm FIR}$ is the integrated luminosity between 42$\,\mu$m and 122$\,\mu$m) and dust
temperatures (i.e.\ $T_{\rm dust}$) by fitting the available FIR
photometry using the DH SED template library and a standard $\chi^2$
minimisation method.  From the integration of the best-fitting DH SED
template, we infer $L_{\rm FIR}$ to within $\thicksim0.1\,$ dex, even
in cases with only one FIR detection, because \textit{Herschel}
observations probe the peak of the FIR emission of galaxies
\citep{elbaz_2011, nordon_2012}.  Naturally, we compared these $L_{\rm
  FIR}$ estimates to the obscured SFR (i.e.\ SFR$_{\rm IR}$) expected
from our `ladder of SFR indicators'.  We reject
SFR--$M_{\ast}$--$\,z$ bins in which these two independent SFR$_{\rm
  IR}$ estimates are not consistent within $0.3\,$dex.  Such
discrepancies are only observed in few SFR--$M_{\ast}$--$\,z$ bins
with stacked FIR flux densities with low significance,
S/N$\,\thicksim\,$$3$.

From the best-fitting DH SED template we also estimated $T_{\rm dust}$
using the pairing between dust temperature and DH templates
established in \citet{magnelli_2014}.  The reliability of these dust
temperature estimates depends on the number of FIR data points
available and on whether those data points encompass the peak of the
FIR emission.  As in \citet{magnelli_2014}, we considered as reliable
only the dust temperatures inferred from at least three FIR data
points encompassing the peak of the FIR emission and with $\chi^2_{\rm
  reduced}$$\,<\,$$3$.
\begin{figure*}
\center
\includegraphics[width=9.15cm]{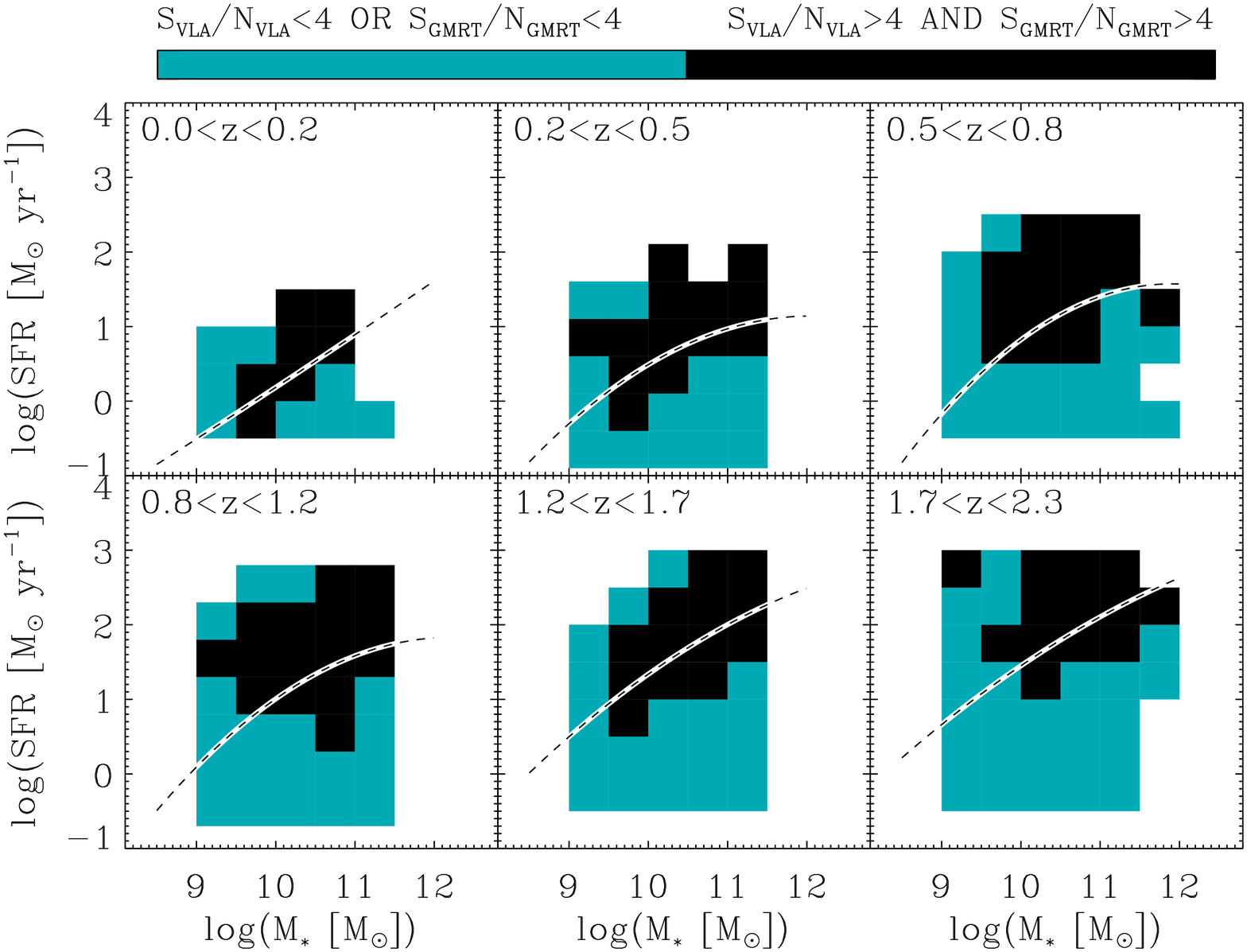}
\includegraphics[width=9.15cm]{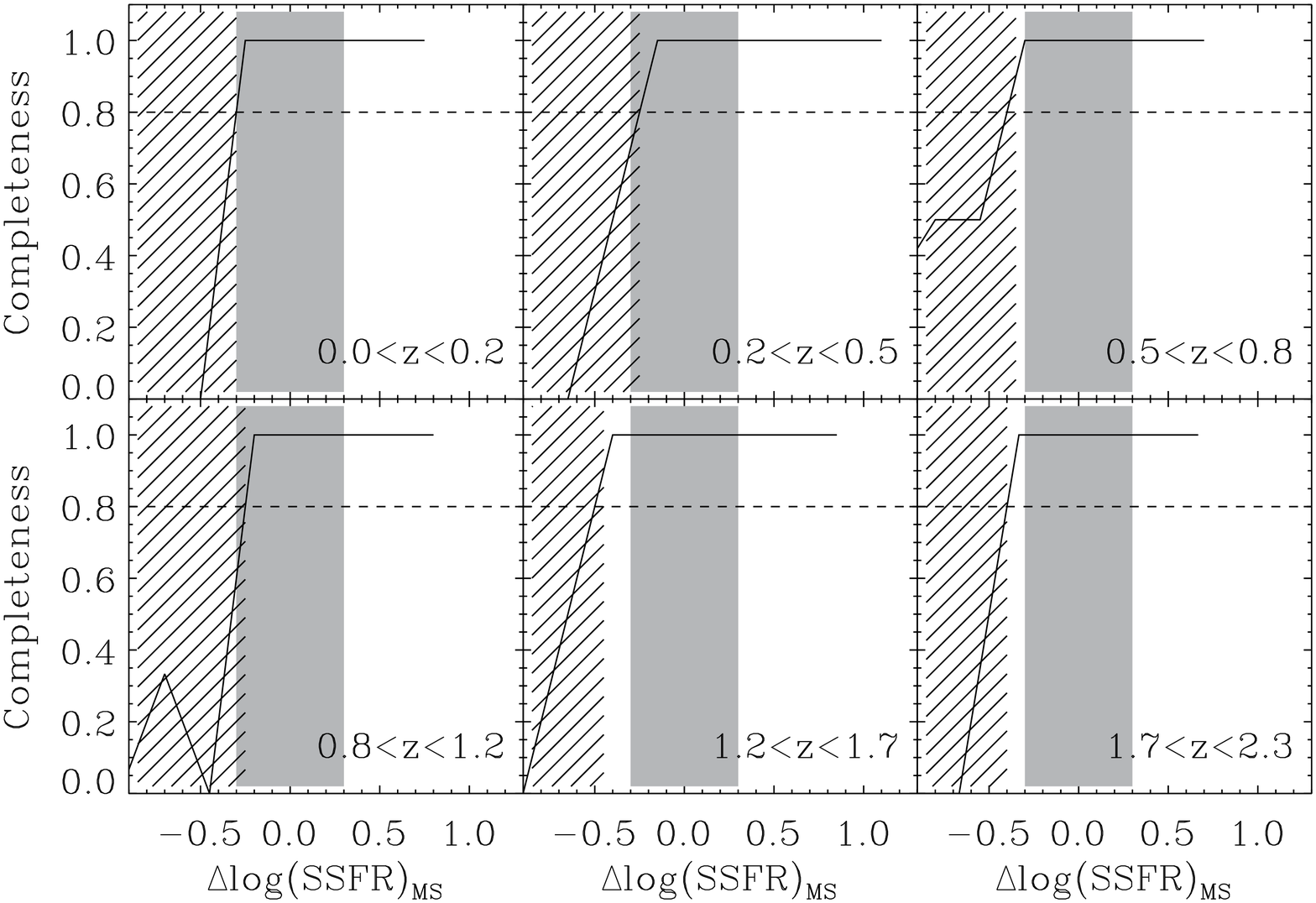}
\caption{ \label{fig:completeness index} (\textit{left})
  SFR$-M_{\ast}$ bins with accurate $\alpha^{\rm 1.4\,GHz}_{\rm
    610\,MHz}$ estimates from our stacking analysis.  Estimates are
  considered accurate if $S_{\rm 1.4\,GHz}/N_{\rm
    1.4\,GHz}$$\,>\,$$4$ (i.e.\ signal, $S$, over noise, $N$, ratio) and $S_{\rm 610\,MHz}/N_{\rm
    610\,MHz}$$\,>\,$$4$.  Short-dashed lines on a white background
  show the MS of star formation.  (\textit{right}) Fraction of
  SFR$-M_{\ast}$ bins with $M_{\ast}$$\,>\,$$10^{10}\,$M$_{\odot}$ and
  with accurate $\alpha^{\rm 1.4\,GHz}_{\rm 610\,MHz}$ estimates as
  function of their $\Delta$log$({\rm SSFR})_{\rm MS}$.  Horizontal
  dashed lines represent the 80\% completeness limits.  Hatched areas
  represent the regions of parameter space affected by incompleteness,
  i.e.\ where less than 80\% of our SFR$-M_{\ast}$ bins have accurate
  $\alpha^{\rm 1.4\,GHz}_{\rm 610\,MHz}$ estimates.  Shaded regions
  show the location and dispersion of the MS of star formation.  }
\end{figure*}

\begin{figure*}
\center
\includegraphics[width=9.15cm]{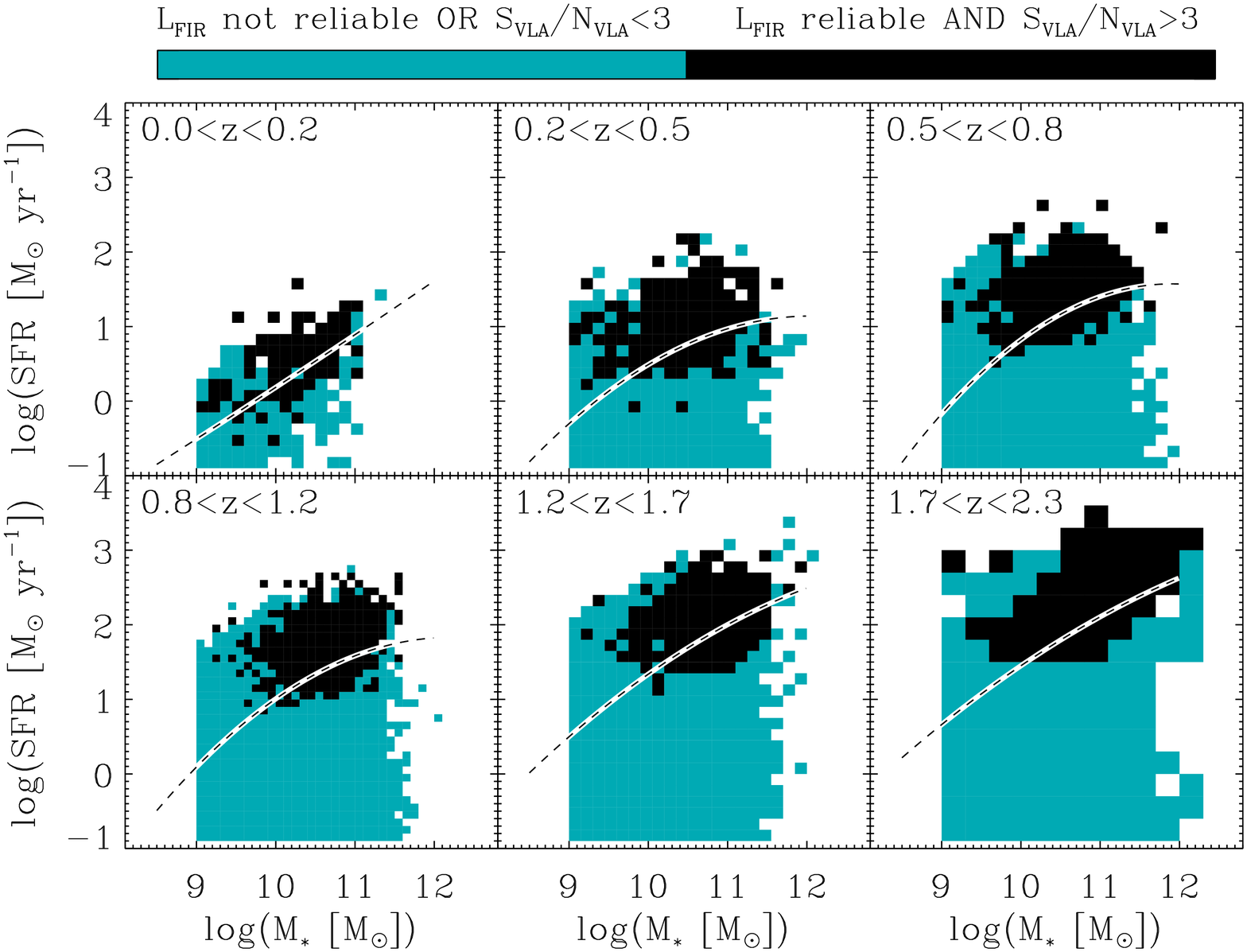}
\includegraphics[width=9.15cm]{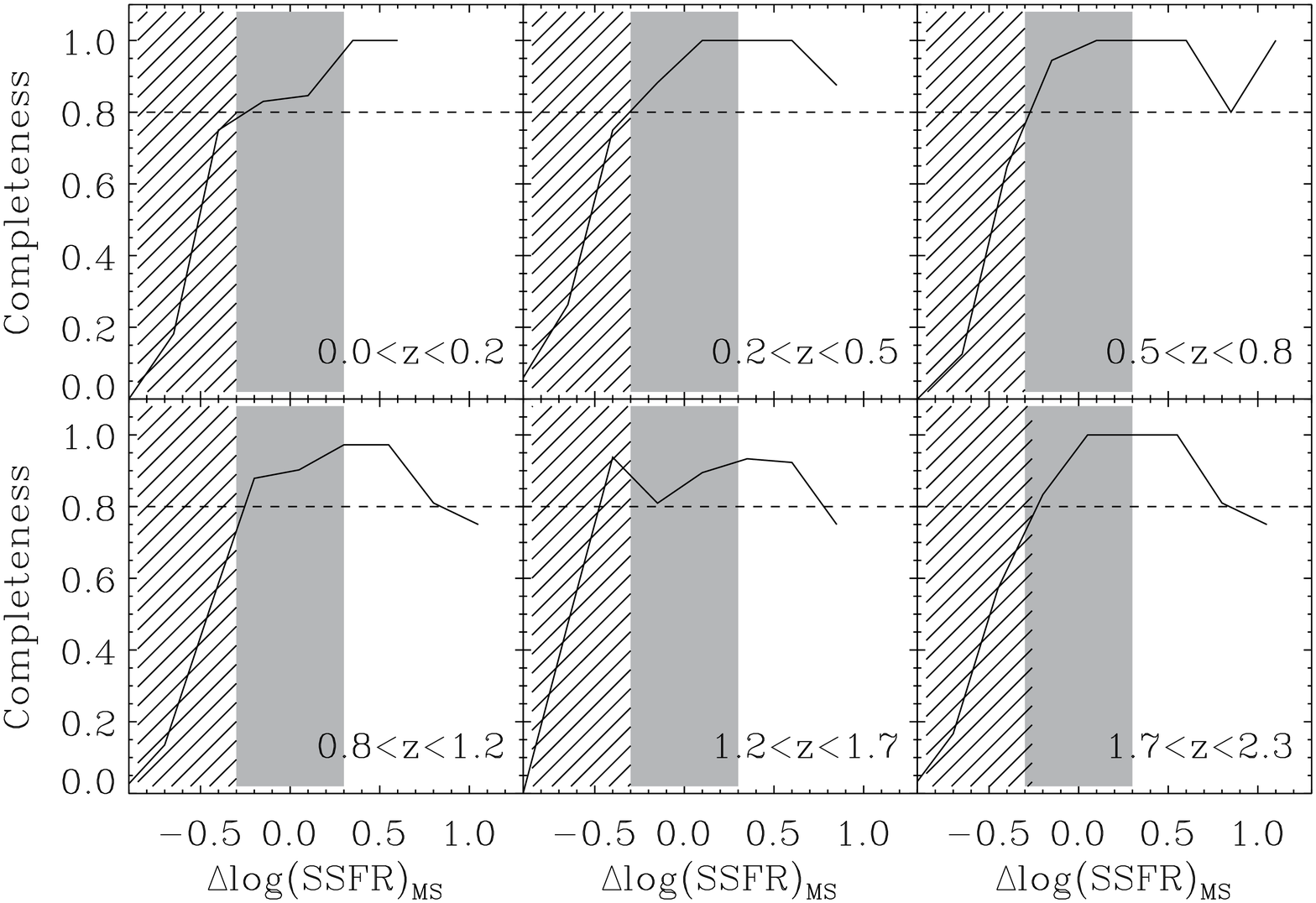}
\caption{ \label{fig:completeness q} (\textit{left}) SFR$-M_{\ast}$
  bins with accurate $q_{\rm FIR}$ estimates from our stacking
  analysis.  Estimates are considered accurate if $S_{\rm
    1.4\,GHz}/N_{\rm 1.4\,GHz}$$\,>\,$$3$ (i.e.\ signal, $S$, over noise, $N$, ratio) and $L_{\rm FIR}$ is
  reliable (see Section~\ref{subsec:FIR stacking}).  Short-dashed lines
  on a white background show the MS of star formation.
  (\textit{right}) Fraction of SFR$-M_{\ast}$ bins with
  $M_{\ast}$$\,>\,$$10^{10}\,$M$_{\odot}$ and with accurate $q_{\rm
    FIR}$ estimates as function of their $\Delta$log$({\rm SSFR})_{\rm
    MS}$.  Horizontal dashed lines represent the 80\% completeness
  limits.  Hatched areas represent the regions of parameter space
  affected by incompleteness.  Shaded regions show the location and
  dispersion of the MS of star formation.  }
\end{figure*}

 \subsection{Determination of the radio properties through stacking\label{subsec:radio stacking}}

 To estimate the radio properties (i.e.\ $S_{\rm 1.4\,GHz}$ and
 $S_{\rm 610\,MHz}$) of a given galaxy population, we stacked their
 VLA 1.4-GHz and GMRT 610-MHz observations.  This stacking analysis is
 very similar to that employed for the \textit{Herschel} observations.
 However, the \textit{Herschel} data are very homogeneous among
 fields, except for the noise level, whereas the radio data differ
 between fields in properties such as beam shape.  Hence, one cannot
 stack together sources regardless of their position on the sky.
 Instead, one can stack together sources of the same field, measure their stacked flux densities and associated errors, and finally combine information from different fields using a weighted mean. For
 a given galaxy population (i.e.\ a given SFR--$M_{\ast}$--$\,z$ bin),
 for each field and each radio band ($1.4\,$GHz and $610\,$MHz), we thus
 proceeded as follows.  We stacked all sources (regardless of whether
 they are detected or not) situated in a given field using the
 original VLA (or GMRT) images.  We measured the radio stacked flux
 density of this galaxy population in this field by fitting their
 median radio stacked stamp with a 2D Gaussian function.  Use of the
 median avoids the biasing influence of moderately radio-loud AGNs
 \citep[e.g.][]{delmoro_2013} that have not yet been removed from the
 source list via X-ray emission.  Because the S/N of our typical radio
 stacked stamp was poor, in these fits we only left the normalisation
 of the 2D Gaussian function as a free parameter (i.e.\ $S_{\rm
   Gaussian}$).  The position, minor and major axes (i.e.\
 [$a\,$,$\,b$]), and position angle of this 2D Gaussian function were
 fixed to the values found when fitting the high S/N radio stacked
 stamp of all galaxies within the current field, redshift bin and with
 $M_{\ast}$$\,>\,$$10^{10}\,$M$_{\odot}$ and $\Delta$log$({\rm
   SSFR})_{\rm MS}$$\,>\,$$-0.3$. The radio stacked flux density of
 this SFR--$M_{\ast}$--$\,z$ bin in this field (i.e.\ $S_{\rm
   radio}^{i}$) was then given by
\begin{equation}
S_{\rm radio}^{i}= \frac{S_{\rm Gaussian}\times a\times b}{a_{\rm norm}\times b_{\rm norm}},
\end{equation}
where $a_{\rm norm}$ and $b_{\rm norm}$ are the minor and major axes
of the radio beam in the original VLA (or GMRT) images.
For the VLA observations, where the spatial resolution is relatively high and thus beam smearing by astrometric uncertainties in the stacked samples can be significant, $S_{\rm radio}$ were $\thicksim$$1.9$ times higher than $S_{\rm Gaussian}$.
Despite astrometric uncertainties, galaxy sizes marginally resolved at the $\thicksim$$\,1.5\arcsec$ resolution of the VLA observations might also explain part of the difference between $S_{\rm radio}$ and $S_{\rm Gaussian}$. The appropriate uncertainty on this radio flux density was obtained from a
bootstrap analysis.  Finally, we combined the radio stacked flux
densities of a given SFR--$M_{\ast}$--$\,z$ bin from different fields
(i.e.\ GOODS-N, GOODS-S, ECDFS and COSMOS) using a weighted mean.
\begin{figure*}
\center
\includegraphics[width=14.9cm]{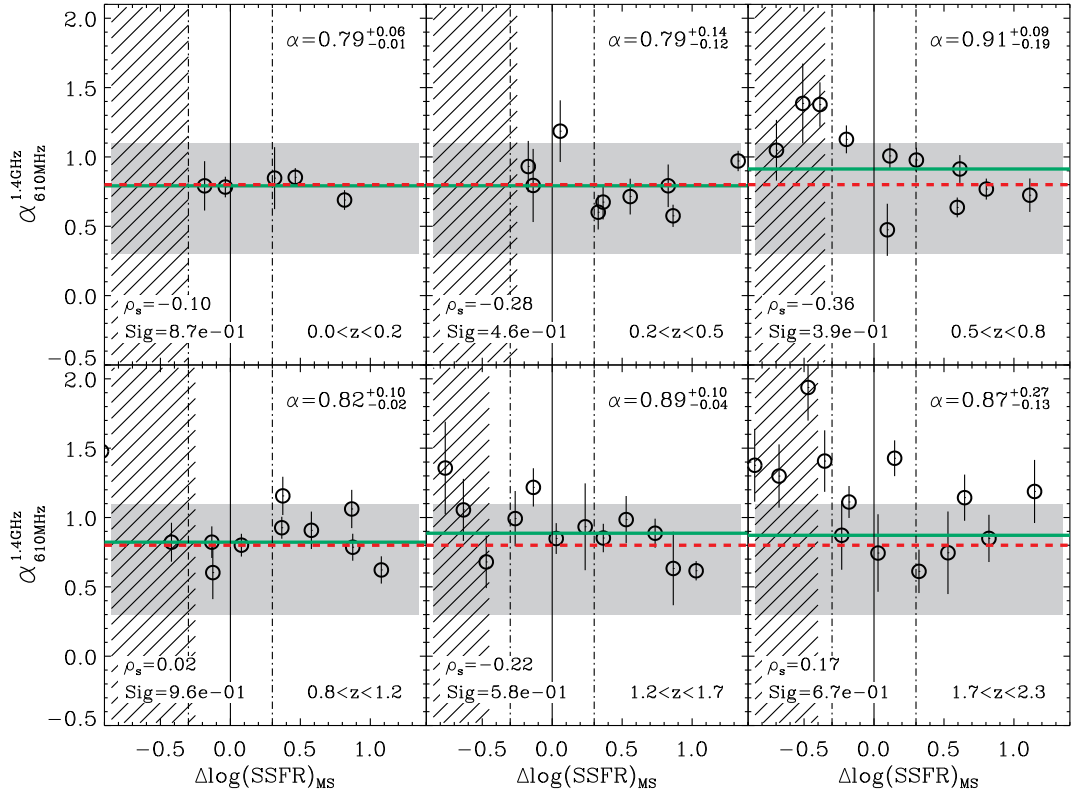}
\caption{ \label{fig:index} Radio spectral index (i.e.\ $\alpha^{\rm
    1.4\,GHz}_{\rm 610\,MHz}$) of galaxies as a function of
  $\Delta$log$({\rm SSFR})_{\rm MS}$, as derived from our stacking
  analysis.  Hatched areas represent the regions of parameter space
  affected by incompleteness (see text and Fig.~\ref{fig:completeness
    index}).  In each panel, we give the median value (see also the
  green lines), the Spearman rank correlation ($\rho_{\rm s}$) and the
  null hypothesis probability (Sig.) derived from data points in the
  region of parameter space not affected by incompleteness.  Red
  dashed lines correspond to the \textit{canonical} radio spectral
  index of 0.8 observed in local star-forming galaxies
  \citep{condon_1992} and high-redshift SMGs \citep{ibar_2010}.  Shaded regions show the range of $\alpha^{\rm
    1.4\,GHz}_{\rm 610\,MHz}$ values (i.e.\ $0.7\pm0.4$) observed by
  \citet{ibar_2009} in a population of sub-mJy radio galaxies.
  Vertical solid and dot-dashed lines show the localisation and width
  of the MS of star formation.  In the range of redshift and
  $\Delta$log$({\rm SSFR})_{\rm MS}$ probed here, $\alpha^{\rm
    1.4\,GHz}_{\rm 610\,MHz}$ does not significantly deviate from its
  canonical value, 0.8.  }
\end{figure*}

From their stacked 1.4-GHz flux densities, we derived the rest-frame
1.4-GHz luminosity (i.e.\ $L_{\rm 1.4\,GHz}$) for our
SFR--$M_{\ast}$--$\,z$ bins.  For that, we $k$-corrected their stacked
1.4-GHz flux density assuming a radio spectral index of
$\alpha$$\,=\,$$0.8$ \citep[e.g.][]{condon_1992,ibar_2009,ibar_2010}.  This
value is perfectly in line with the radio spectral index observed here
(i.e.\ $\alpha^{\rm 1.4\,GHz}_{\rm 610\,MHz}$) over a broad range of
redshift and $\Delta$log$({\rm SSFR})_{\rm MS}$ (see
Section~\ref{subsec:spectral index}).

\subsection{The SFR--$M_{\ast}$--$\,z$ parameter space\label{subsec:completeness}}

Before looking at the evolution of the FRC and radio spectral index in
the SFR--$M_{\ast}$--$\,z$ parameter space, we need to ensure that our
ability to make accurate $q_{\rm FIR}$ and $\alpha^{\rm 1.4\,GHz}_{\rm
  610\,MHz}$ measurements does not introduce significant
incompleteness in any particular regions of the SFR$-M_{\ast}$ plane.
The left panels of Figs~\ref{fig:completeness index} and
\ref{fig:completeness q} present the regions of the
SFR--$M_{\ast}$--$\,z$ parameter space with accurate $\alpha^{\rm
  1.4\,GHz}_{\rm 610\,MHz}$ and $q_{\rm FIR}$ estimates from our
stacking analysis, respectively.  
The sampling of the SFR-$M_{\ast}$ plane is made with larger SFR-$M_{\ast}$ bins for $\alpha^{\rm 1.4\,GHz}_{\rm 610\,MHz}$ than for $q_{\rm FIR}$. 
  This is due to the fact that the GMRT observations covered a sky area of only $0.3\,$deg$^2$ (Sect.~\ref{subsec:GMRT}).
  This limited dataset forces us to enlarge the size of our SFR-$M_{\ast}$ bins in order to increase the number of stacked sources per bin and thus improve the noise in our stacked stamps ($\sigma_{\rm stack}$$\,\propto\,$$\sqrt{N}$).
Our $\alpha^{\rm 1.4\,GHz}_{\rm
  610\,MHz}$ estimates are considered as accurate only if $S_{\rm
  1.4\,GHz}/N_{\rm 1.4\,GHz}$$\,>\,$$4$ and $S_{\rm 610\,MHz}/N_{\rm
  610\,MHz}$$\,>\,$$4$.  Our $q_{\rm FIR}$ estimates are considered as
accurate only if $S_{\rm 1.4\,GHz}/N_{\rm 1.4\,GHz}$$\,>\,$$3$ and
$L_{\rm FIR}$ is reliable (see Section~\ref{subsec:FIR stacking}).  In
each of our redshift bins, our stacking analysis allows us to obtain
accurate $\alpha^{\rm 1.4\,GHz}_{\rm 610\,MHz}$ and $q_{\rm FIR}$
estimates for almost all MS and above-MS galaxies with
$M_{\ast}$$\,>\,$$10^{10}\,$M$_{\odot}$.

The right panels of Figs~\ref{fig:completeness index} and
\ref{fig:completeness q} illustrate our ability to study the
variations of $\alpha^{\rm 1.4\,GHz}_{\rm 610\,MHz}$ and $q_{\rm FIR}$
for galaxies with $M_{\ast}$$\,>\,$$10^{10}\,$M$_{\sun}$,
respectively.  In these figures we show the fraction of SFR$-M_{\ast}$
bins with reliable $\alpha^{\rm 1.4\,GHz}_{\rm 610\,MHz}$ or $q_{\rm
  FIR}$ estimates as a function of their $\Delta$log$({\rm SSFR})_{\rm
  MS}$.  In the rest of the paper, we consider that
$\alpha^{\rm 1.4\,GHz}_{\rm 610\,MHz}$ or $q_{\rm FIR}$ in a given
$\Delta$log$({\rm SSFR})_{\rm MS}$ bin is fully constrained only if
the completeness in this bin is $\gtrsim$80\%.  In each redshift bin,
our stacking analysis allows us to fully constrain $\alpha^{\rm
  1.4\,GHz}_{\rm 610\,MHz}$ and $q_{\rm FIR}$ in galaxies with
$\Delta$log$({\rm SSFR})_{\rm MS}$$\,>\,$$-0.3$.

From this analysis, we conclude that our stacking analysis provides us
with a complete view on the evolution of $\alpha^{\rm 1.4\,GHz}_{\rm
  610\,MHz}$ and $q_{\rm FIR}$ up to $z$$\,\thicksim\,$$2$ in
star-forming galaxies with $M_{\ast}$$\,>\,$$10^{10}\,$M$_{\odot}$ and
$\Delta$log$({\rm SSFR})_{\rm MS}$$\,>\,$$-0.3$.

\section{Results\label{sec:results}}

\subsection{The radio spectral index, $\alpha^{\rm 1.4\,GHz}_{\rm 610\,MHz}$\label{subsec:spectral index}}

The radio spectral index of each SFR--$M_{\ast}$--$\,z$ bin was
inferred using their VLA and GMRT stacked flux densities, i.e.\
$S_{\rm 1.4\,GHz}$ and $S_{\rm 610\,MHz}$, respectively.  Assuming
that the radio spectrum follows a power law form,
$S_{\nu}$$\,\propto\,$$\nu^{-\alpha}$, the radio spectral index is
given by,
\begin{equation}
\label{eq.radio spectral index}
\alpha^{\rm 1.4\,GHz}_{\rm 610\,MHz}=\frac{{\rm log}(S_{\rm 610\,MHz}/S_{\rm 1.4\,GHz})}{{\rm log}(1400/610)}.
\end{equation}
$S_{\rm 1.4\,GHz}$ and $S_{\rm 610\,MHz}$ are \textit{observed} flux densities. Therefore, at different redshift, $\alpha^{\rm 1.4\,GHz}_{\rm 610\,MHz}$ correspond to radio spectral indices at different rest-frame frequencies. This has to be taken into account when interpreting our results (see Sect. \ref{subsubsect:nature radio}).
In the following, we discuss only accurate $\alpha^{\rm 1.4\,GHz}_{\rm
  610\,MHz}$ estimates.
We remind the reader that constraints on the radio spectral index of star-forming galaxies are obtained from a somewhat limited 
 galaxy sample, covering a sky area of $\thicksim\,$$0.3\,$deg$^2$, as GMRT observations of the COSMOS fields are not available.

The Fig.~\ref{fig:index} presents the evolution of $\alpha^{\rm
  1.4\,GHz}_{\rm 610\,MHz}$ as a function of $\Delta$log$({\rm
  SSFR})_{\rm MS}$ up to $z$$\,\thicksim\,$$2$.  We find no strong
correlation between $\alpha^{\rm 1.4\,GHz}_{\rm 610\,MHz}$ and
$\Delta$log$({\rm SSFR})_{\rm MS}$ in any redshift bin (i.e.\
$|\,\rho_{\rm s}|$$\,\lesssim\,0.6$) and the null hypothesis of
uncorrelated data cannot be rejected with high significance (i.e.\
Sig.$\,>\,$5\%).  In addition, the median $\alpha^{\rm 1.4\,GHz}_{\rm
  610\,MHz}$ does not deviate significantly from the canonical value, 0.8
\citep[e.g.][]{condon_1992}, in any of our redshift bins.  From
these findings we conclude that most star-forming galaxies with
$M_{\ast}$$\,>\,$$10^{10}\,$M$_{\odot}$ and across
$0$$\,<\,$$z$$\,<\,$$2.3$ have on average a radio spectral index consistent with
0.8.  However, due to the relatively low number of data
points and to the large dispersion on $\alpha^{\rm 1.4\,GHz}_{\rm
  610\,MHz}$, we cannot confidently rule out the presence of a
negative, but weak, $\alpha^{\rm 1.4\,GHz}_{\rm
  610\,MHz}\,$--$\,\Delta$log$({\rm SSFR})_{\rm MS}$ correlation
($\rho_{\rm s}$$\,<\,$$0$).  Such a trend would echo (but not match)
the results of \citet[see also Clemens et al.
\citeyear{clemens_2008}]{condon_1991} who found that the most extreme
local starbursts have flatter radio spectra (i.e.\ $\thicksim\,$$0.5$)
than local normal star-forming galaxies (i.e.\ $\thicksim\,$$0.8$).
\citet{condon_1991} and \citet{clemens_2008} attribute these flat
radio spectra to free-free absorption in dense nuclear starbursts.

\begin{figure}
\center
\includegraphics[width=9.cm]{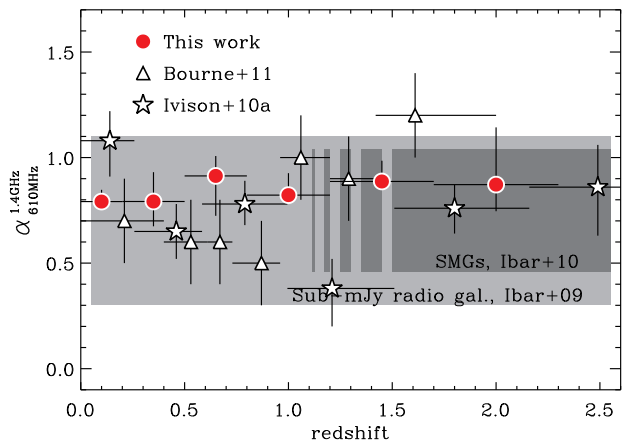}
\caption{ \label{fig:index z} Evolution of the radio spectral index,
  $\alpha^{\rm 1.4\,GHz}_{\rm 610\,MHz}$, with redshift, as inferred
  from our stacking analysis.  Red circles and error bars correspond
  to the median and interquartile range observed in our study in
  regions of the parameter space not affected by incompleteness (see
  values and green lines reported in Fig.~\ref{fig:index}).  Stars
  show results from \citet{ivison_2010a}, while triangles show results
  from \citet{bourne_2011}.  Both studies are based on a stacking
  analysis, but \citet{ivison_2010a} applied it to a sample of
  24-$\mu$m-selected galaxies while \citet{bourne_2011} applied
  it to a stellar-mass-selected galaxy sample.  The light
  grey region shows the range of $\alpha^{\rm 1.4\,GHz}_{\rm
    610\,MHz}$ values (i.e.\ $0.7\pm0.4$) observed by
  \citet{ibar_2009} in a population of sub-mJy radio galaxies.  The
  dark grey region presents the range of $\alpha^{\rm 1.4\,GHz}_{\rm
    610\,MHz}$ values (i.e.\ $0.75\pm0.29$) observed in a population
  of SMGs at $z\thicksim2$ \citep{ibar_2010}. Note that the ranges of values observed in \citet{ibar_2009,ibar_2010} correspond to the intrinsic dispersion observed in these galaxy populations and not to measurement errors on the mean radio spectral indices.}
\end{figure}

The Fig.~\ref{fig:index z} shows that our results agree well with the
range of radio spectral index observed in a large sample of sub-mJy
radio galaxies \citep{ibar_2009} and in a population of
$z$$\,\thicksim\,$$2$ SMGs \citep[][see also
\citealt{thomson_2014}]{ibar_2010}.  These agreements are re-assuring
because results from \citet{ibar_2009,ibar_2010} and \citet{thomson_2014} were
based on galaxies individually detected in the VLA and GMRT images.

Our results are also in line with those of \citet{bourne_2011} and
\citet{ivison_2010a}.  Both studies are based on a stacking analysis,
but \citet{ivison_2010a} applied it to a sample of 24-$\mu$m-selected (i.e.\ SFR-selected)
galaxies while \citet{bourne_2011} applied it to a
stellar-mass-selected galaxy sample (including both star-forming and quiescent galaxies).  The consistencies
observed between studies with different selection functions re-inforce
our conclusion that $\alpha^{\rm 1.4\,GHz}_{\rm 610\,MHz}$ does not
significantly evolve across $0$$\,<\,$$z$$\,<\,$$2.3$.  We note that
results from our study and those from \citet{ivison_2010a} and
\citet{bourne_2011} are, however, not strictly independent since
they are based on the same VLA and similar GMRT observations of ECDFS.
Nevertheless, we believe that the consistencies found here are
noteworthy because those studies differ in many other aspects: our
stacking analysis includes observations from the GOODS-N field; our
GMRT observations of ECDFS are deeper; our stellar-mass-selected
sample is built using different optical-to-near-IR multi-wavelength
catalogues, applying different methods to infer photometric redshifts
and stellar masses.

The absence of \textit{significant} evolution in $\alpha^{\rm
  1.4\,GHz}_{\rm 610\,MHz}$ with $z$ and $\Delta$log$({\rm SSFR})_{\rm
  MS}$ is an important result for our forthcoming study of the FRC.
Indeed, to $k$-correct our stacked 1.4-GHz flux densities into
rest-frame 1.4-GHz radio luminosities, we assumed that the radio
spectral index of \textit{all} galaxies across
$0$$\,<\,$$z$$\,<\,$$2.3$ was equal to 0.8 (see
Section~\ref{subsec:radio stacking}).  Significant deviations from
this canonical value would have introduced artificial evolution of
the FRC in the SFR--$M_{\ast}$--$\,z$ parameter space ($\Delta q_{\rm FIR}$$\,=\,$$-\Delta \alpha\,$$\times\,$log(1+$z$)).

The physical implications of the absence of \textit{significant} evolution of $\alpha^{\rm
  1.4\,GHz}_{\rm 610\,MHz}$ with $z$ and $\Delta$log$({\rm SSFR})_{\rm
  MS}$ are discussed in Sect.~\ref{sec:discussion}.

\begin{figure*}
\center
\includegraphics[width=13.8cm]{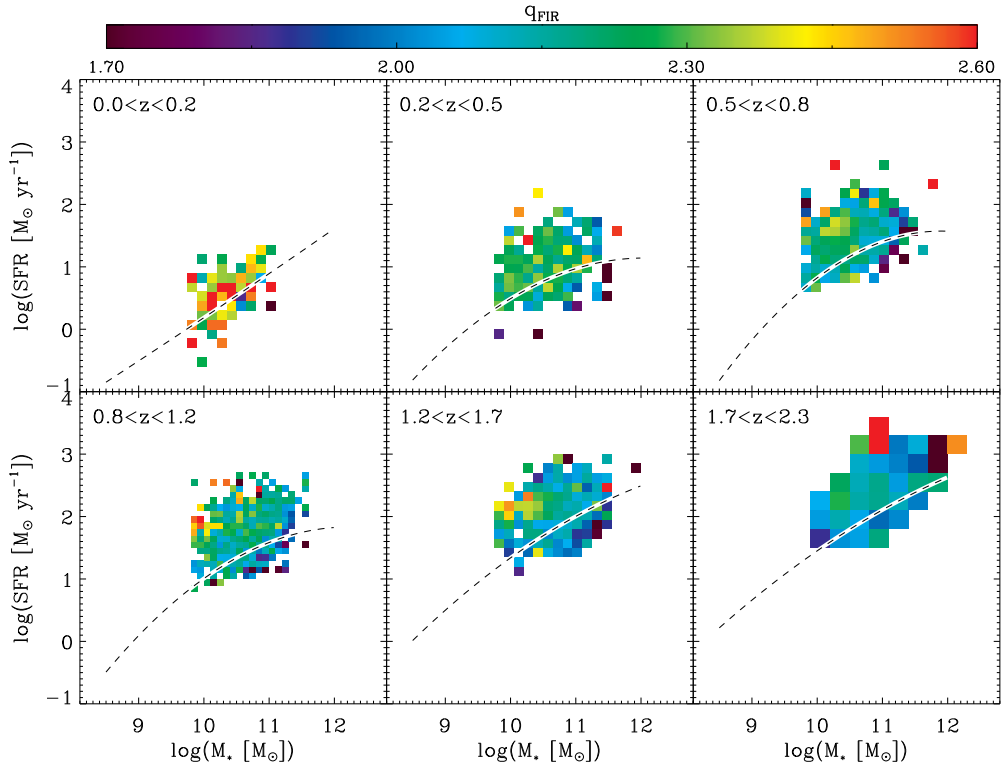}
\caption{ \label{fig:q} Evolution of the mean $L_{\rm FIR}\,$-- to
  --$\,L_{\rm 1.4\,GHz}$ ratio, i.e.\ $q_{\rm FIR}$ (see
  Eq.~\ref{eq:fir/radio}), of galaxies in the SFR$-M_{\ast}$ plane, as
  found using our stacking analysis.  Short-dashed lines on a white
  background show the MS of star formation.  }
\end{figure*}

\begin{figure*}
\center
\includegraphics[width=14.9cm]{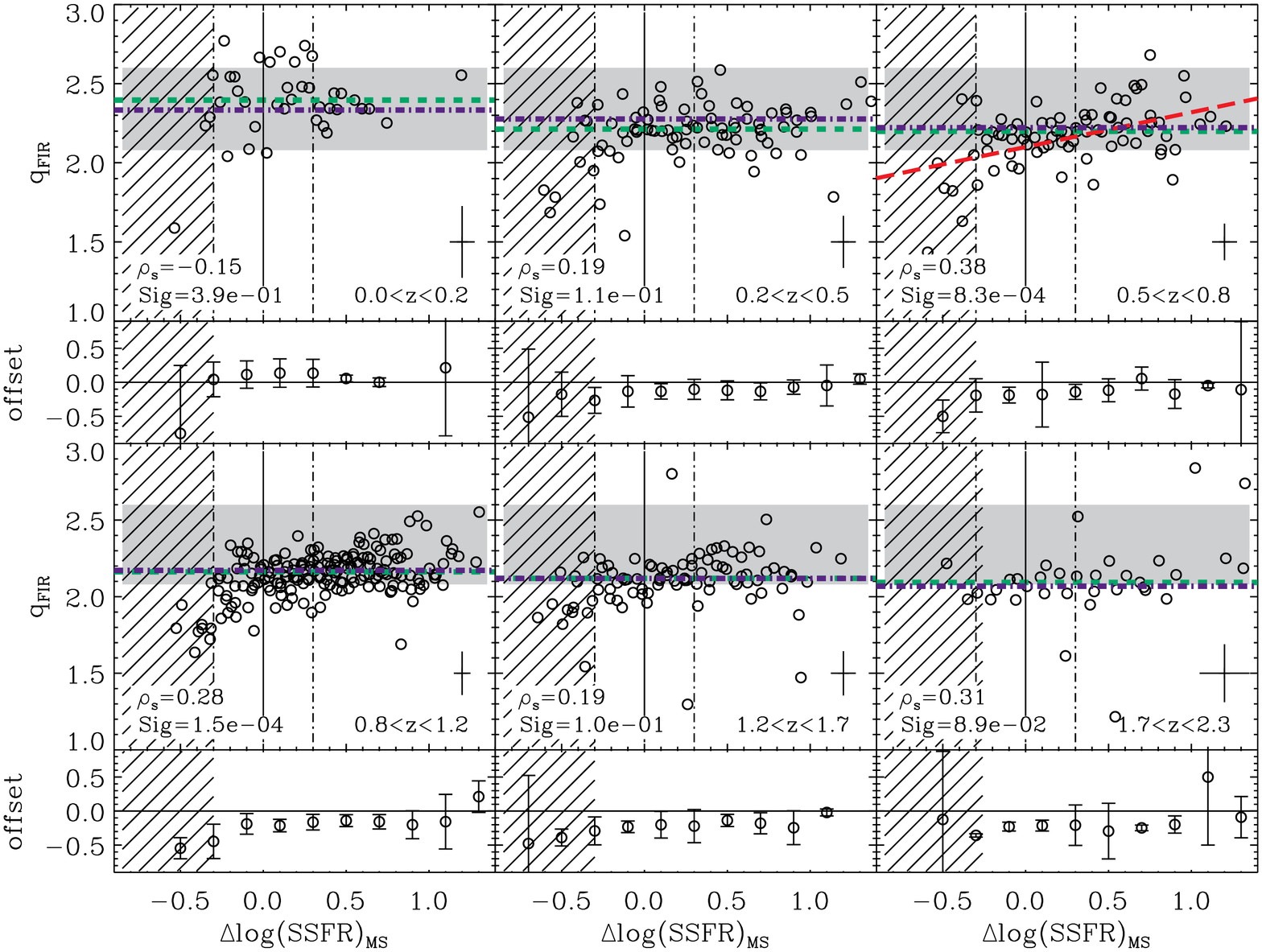}
\caption{ \label{fig:q dssfr} Mean $L_{\rm FIR}\,$-- to --$\,L_{\rm
    1.4\,GHz}$ ratio ($q_{\rm FIR}$) of galaxies as a function of
  $\Delta$log$({\rm SSFR})_{\rm MS}$, as derived from our stacking
  analysis.  Hatched areas represent the regions of parameter space
  affected by incompleteness (see text and Fig.~\ref{fig:completeness
    q}), while the light grey region shows the value of $q_{\rm FIR}$ observed by \citet{yun_2001} in a large sample of local star-forming galaxies, $q_{\rm FIR}({\footnotesize z\thicksim0})$$\,=\,$$2.34\pm0.26$. In each panel, we show the median value (green dashed line),
  give the Spearman rank correlation ($\rho_{\rm s}$) and the null
  hypothesis probability (Sig.) derived from data points in the region
  of parameter space not affected by incompleteness. Blue dot-dashed
  lines represent a redshift evolution of $q_{\rm
    FIR}$$\,=\,$$2.35\times(1+z)^{-0.12}$.  In the lower right part of
  each panel, we give the median uncertainty on our $q_{\rm FIR}$
  estimates.  Vertical solid and dot-dashed lines show the
  localisation and the width of the MS of star formation.  
  In the redshift bin with a statistically significant
 correlation (Sig.$\,<\,$$5\,$$\%$ and $\chi^{2}_{\rm red}[{\rm linear}]$$\,<\,$$\chi^{2}_{\rm red}[{\rm cst}]$, see text), we plot (red long-dashed line) a linear fit, $q_{\rm FIR}= (0.22\pm0.07)\times\Delta{\rm log(SSFR)_{\rm MS}}+(2.10\pm0.05)$.
  The lower panel
  of each redshift bin shows the offset between the median $q_{\rm
    FIR}$ of our data and the local value of $q_{\rm
    FIR}(${\footnotesize $z$$\,=\,$$0$}$)$$\,\approx\,$$2.34$, in bins
  of $0.2\,$dex.}
\end{figure*}

\subsection{The FIR/radio correlation\label{subsec:q}}

Using the rest-frame FIR and 1.4-GHz luminosities estimated from
our stacking analysis, we study the evolution of the FRC in the
SFR--$M_{\ast}$--$\,z$ parameter space.  For that, we use the
parametrisation of the FRC given in \citet[][see also Yun et al.
\citeyear{yun_2001}]{helou_1988},
\begin{equation}
\label{eq:fir/radio}
q_{\rm FIR}={\rm log}\left(\frac{L_{{\rm FIR}}[{\rm W}]}{3.75\times10^{12}}\right)-{\rm log}\left(L_{1.4\,{\rm GHz}}[{\rm W\,Hz^{-1}}]\right),
\end{equation}
where $L_{{\rm FIR}}$ is the integrated FIR luminosity from rest-frame 42 to
122$\,\mu$m and $L_{1.4\,{\rm GHz}}$ is the rest-frame 1.4-GHz radio
luminosity density.  Radio $k$-corrections are inferred assuming
$S_{\nu}$$\,\propto\,$$\nu^{-\alpha}$ and the canonical radio spectral
index of $\alpha$$\,=\,$$0.8$ \citep[see Section~\ref{subsec:spectral
  index};][]{condon_1992}.  In the recent literature, alternative
parametrisations of the FRC have been proposed.  In particular, some
studies have used the infrared luminosity from rest-frame 8 to
1000$\,\mu$m ($L_{\rm IR}$) instead of $L_{{\rm FIR}}$
\citep[e.g.][]{ivison_2010a,ivison_2010b,sargent_2010,bourne_2011}.
Such different definitions have no significant impact because there is
a tight relation between $L_{\rm FIR}$ and $L_{\rm IR}$ of star-forming galaxies.
Over all our SFR--$M_{\ast}$--$\,z$ bins, we indeed found $L_{\rm
  IR}$$\,=\,$$1.91_{-0.05}^{+0.10}$$\,\times\,$$L_{\rm FIR}$.  Thus, to
compare $q_{\rm FIR}$ with $q_{\rm IR}$, we simply corrected these
estimates following $q_{\rm FIR}$$\,=\,$$q_{\rm IR}-{\rm log}(1.91)$.

The Fig.~\ref{fig:q} shows the evolution of $q_{\rm FIR}$ in the
SFR$-M_{\ast}$ plane in different redshift bins. 
In this figure, we do not identify any significant and systematic evolution of $q_{\rm FIR}$ in the SFR-$M_{\ast}$ plane.
In contrast, we notice a clear and systematic decrease in $q_{\rm FIR}$ with redshift.

In Fig.~\ref{fig:q dssfr} we investigate the existence of more subtle evolution of $q_{\rm FIR}$ within the SFR-$M_{\ast}$ plane by plotting the variation of $q_{\rm FIR}$ as a function of $\Delta$log$({\rm SSFR})_{\rm MS}$.
In all our redshift bins, there is a weak ($0.1$$\,\lesssim\,$$|\,\rho_{\rm
  s}|$$\,\lesssim\,$$0.4$) correlation between $q_{\rm FIR}$ and
$\Delta$log$({\rm SSFR})_{\rm MS}$, though the null
hypothesis of uncorrelated data can only be rejected with high significance (i.e.\
Sig.$\,<\,$5\%) in two of these redshift bins (i.e.\
$0.5$$\,<\,$$z$$\,<\,$$0.8$ and $0.8$$\,<\,$$z$$\,<\,$$1.2$).
In addition, at high $\Delta$log$({\rm SSFR})_{\rm MS}$, the dispersion on $q_{\rm FIR}$ seems to increase.

To study further the statistical significance of a weak $q_{\rm FIR}\,$--$\,\Delta$log$({\rm SSFR})_{\rm MS}$ correlation, we fit this relation with a constant and with a linear function using a Monte-Carlo approach taking into account errors both on $q_{\rm FIR}\,$ and $\,\Delta$log$({\rm SSFR})_{\rm MS}$.
Data points used in these fits are restricted to those situated in regions of parameter space not affected by incompleteness.
For each of our $1\,000$ Monte-Carlo realisations, we adopt new values of $q_{\rm FIR}$ and $\Delta$log$({\rm SSFR})_{\rm MS}$ selected into a Gaussian distribution centred on their original values and with a dispersion given by their measurement errors.
To ensure that our fits are not dominated by few data points, for each Monte-Carlo realisation we resample the observed dataset, keeping its original size but randomly selecting its data points with replacement.
We then fit each Monte-Carlo realisation with a constant and with a linear function.
Finally, we study the mean value and dispersion of each fitting parameter across our $1\,000$ Monte-Carlo realisations.

In all but the $0.5$$\,<\,$$z$$\,<\,$$0.8$ redshift bin, the constant model has reduced $\chi^{2}$ values lower than those of the linear model.
In addition, in these redshift bins, the linear model has slopes (i.e.\ $\Delta[q_{\rm FIR}]$/$\Delta$[$\Delta$log$({\rm SSFR})_{\rm MS}]$) consistent, within 1$\sigma$, with zero. 
This suggests that there is no significant $q_{\rm FIR}\,$--$\,\Delta$log$({\rm SSFR})_{\rm MS}$ correlation in these redshift bins.
In contrast, at $0.5$$\,<\,$$z$$\,<\,$$0.8$, the linear model is statistically slightly better than the constant model (i.e.\ $\chi^{2}_{\rm red}[{\rm linear}]=7.0$ and $\chi^{2}_{\rm red}[{\rm cst}]=7.5$) and its slope is different to zero at the $\thicksim\,$3$\sigma$ level (i.e.\ $0.22\pm0.07$).
This suggests a positive but weak $q_{\rm FIR}\,$--$\,\Delta$log$({\rm SSFR})_{\rm MS}$ correlation\footnote{this positive correlation will not be erased but rather enhanced where a weak negative $\alpha^{\rm 1.4\,GHz}_{\rm 610\,MHz}\,$--$\,\Delta$log$({\rm SSFR})_{\rm MS}$ correlation exists (see Section~\ref{subsec:spectral index}).} at $0.5$$\,<\,$$z$$\,<\,$$0.8$.
We note that it is also in this redshift bin that our measurement errors on $q_{\rm FIR}$ are the lowest.
This could explain why an intrinsically weak $q_{\rm FIR}\,$--$\,\Delta$log$({\rm SSFR})_{\rm MS}$ correlation can only be statistically significant at $0.5$$\,<\,$$z$$\,<\,$$0.8$.

This Monte-Carlo approach demonstrates that the existence of a $q_{\rm FIR}\,$--$\,\Delta$log$({\rm SSFR})_{\rm MS}$ correlation is yet difficult to assess from our dataset.
If there exists a $q_{\rm FIR}\,$--$\,\Delta$log$({\rm SSFR})_{\rm MS}$ correlation, it is intrinsically weak.
Unfortunately, our observations are not good enough to firmly reveal or rule out such weak correlation in all our redshift bins.
Thus, we conservatively conclude that there is no \textit{significant} $q_{\rm FIR}\,$--$\,\Delta$log$({\rm SSFR})_{\rm MS}$ correlation across $0$$\,<\,$$z$$\,<\,$$2.3$, though the presence of a weak, positive trend, as observed in one of our redshift bin ($0.5$$\,<\,$$z$$\,<\,$$0.8$), cannot be firmly ruled out using our dataset.
Note that debates on the existence of a weak $q_{\rm FIR}\,$--$\,\Delta$log$({\rm SSFR})_{\rm MS}$ correlation also prevail in the local Universe.
\citet{condon_1991} found that the most extreme local starbursts (i.e.\ $\Delta$log$({\rm SSFR})_{\rm MS}$$\,\thicksim\,$$1$) have higher $q_{\rm FIR}$ and larger dispersions of $q_{\rm FIR}$ than normal star-forming galaxies (i.e.\ $\Delta$log$({\rm SSFR})_{\rm MS}$$\,\thicksim\,$$0$).
In contrast, \citet{helou_1985} and \citet{yun_2001} do not report any statistically significant increase in $q_{\rm FIR}$ in extreme starbursts, though they also found a larger dispersion in $q_{\rm FIR}$ for this population.\\

\begin{figure}
\center
\includegraphics[width=9.0cm]{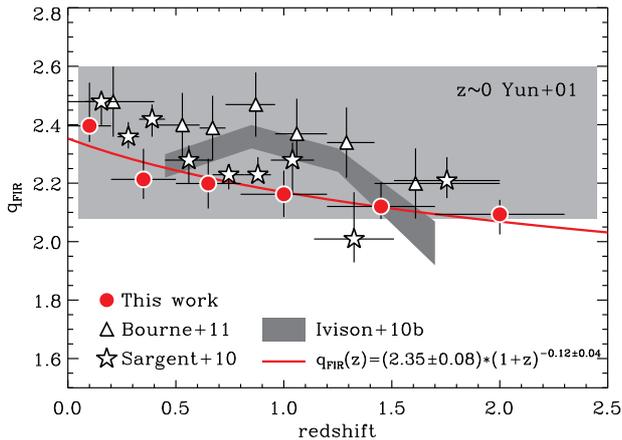}
\caption{ \label{fig:q z} Evolution of the $L_{\rm FIR}\,$-- to
  --$\,L_{\rm 1.4\,GHz}$ ratio, $q_{\rm FIR}$, with redshift, as
  inferred from our stacking analysis.  Red circles and error bars
  correspond to the median and interquartile range observed in our
  study in regions of the parameter space not affected by
  incompleteness (green dashed lines in Fig.~\ref{fig:q dssfr}).  The
  red line corresponds to a redshift-dependent fit to our data points,
  $q_{\rm FIR}(z)$$\,=\,$$2.35\times(1+z)^{-0.12}$.  The light grey
  region shows the value of $q_{\rm FIR}$ observed by \citet{yun_2001}
  in a large sample of local star-forming galaxies, $q_{\rm
    FIR}({\footnotesize z\thicksim0})$$\,=\,$$2.34\pm0.26$.  This \textit{local} measurement is displayed over the entire range of redshift to highlight any possible redshift evolution of $q_{\rm FIR}$. The dark
  grey region presents results from \citet{ivison_2010b} using
  \textit{Herschel} observations on a $L_{\rm IR}$-matched sample.
  Empty triangles show results from \citet{bourne_2011}, as inferred
  using a stacking analysis on a stellar-mass-selected sample of
  galaxies.  Stars present results obtained by \citet{sargent_2010}
  using a FIR/radio-selected sample of star-forming galaxies and
  applying a survival analysis to properly treat flux limits from
  non-detections.  }
\end{figure}

While $q_{\rm FIR}$ does not \textit{significantly} evolve with $\Delta$log$({\rm SSFR})_{\rm MS}$, its median value 
clearly decreases smoothly with redshift.  To study this trend we plot
in Fig.~\ref{fig:q z} the redshift evolution of the median and
interquartile range of $q_{\rm FIR}$, observed in regions of parameter
space not affected by incompleteness.  We find a statistically
significant (Sig.$\,<\,$1\%) redshift evolution of $q_{\rm FIR}$.
This evolution can be parametrised using,
\begin{equation}
\label{eq:evol q}
q_{\rm FIR}(z)=(2.35\pm0.08)\times(1+z)^{-0.12\pm0.04}
\end{equation}
where the $z$$\,=\,$$0$ value of this function agrees perfectly with
local observations, i.e.\ $q_{\rm
  FIR}(z$$\,\thicksim\,$$0)$$\,\approx\,$$2.34\pm0.26$
\citep{yun_2001}.  If we parametrise this redshift
evolution separately for normal ($\Delta$log$({\rm SSFR})_{\rm
  MS}$$\,<\,$$0.75$) and starbursting galaxies ($\Delta$log$({\rm
  SSFR})_{\rm MS}$$\,>\,$$0.75$)\footnote{This definition of
  starbursts is consistent with that of \citet{rodighiero_2011}.}, we
end up with similar solutions, i.e.\ $q_{\rm
  FIR}(z)$$\,=\,$$(2.42\pm0.08)\times(1+z)^{-0.16\pm0.05}$ and $q_{\rm
  FIR}(z)$$\,=\,$$(2.44\pm0.09)\times(1+z)^{-0.13\pm0.09}$,
respectively.  We note that while the evolution of
$q_{\rm FIR}$ with redshift is statistically significant, it is
moderate.  Indeed, at $z$$\,\thicksim\,$$2$, the median value of
$q_{\rm FIR}$ is still within the 1$\sigma$ dispersion of local
observations \citep{yun_2001}.

This redshift evolution of $q_{\rm FIR}$
could not be artificially introduced by a redshift evolution of the
radio spectral index.  Indeed, to create such evolution, the radio
spectral index would need to change from 0.8 to 0.2 between $z$$\,=\,$$0$ and
$z$$\,=\,$$2$, respectively ($\Delta q_{\rm FIR}$$\,=\,$$-\Delta \alpha\,$$\times\,$log(1+$z$)).  Such extreme redshift evolution of the
radio spectral index is not observed in our sample (see
Fig.~\ref{fig:index z} and Section~\ref{subsec:spectral index}).

In Fig.~\ref{fig:q z} we compare our findings with those from the
literature.  In the past decade many papers have discussed this topic. Therefore, instead of presenting an exhaustive comparison, we compare our
findings with the three papers which are, we believe, the most
relevant, i.e.\ those based on relatively complete and well understood
samples and/or valuable FIR/radio datasets.  Firstly, we compare our
results with those of \citet{bourne_2011}.  In that paper, the authors
overcame the selection biases of radio- and/or infrared-selected
samples by using a stellar-mass-selected sample in the ECDFS.  The
infrared and radio properties of their galaxies at a given stellar
mass and redshift were then inferred by stacking \textit{Spitzer} (24,
70 and 160$\,\mu$m) and VLA observations.  This approach is thus very
similar to that employed here.  Secondly, we compare our results with
those of \citet{sargent_2010}.  This paper used a large sample of
infrared- and radio-selected galaxies in the COSMOS field.  To
overcome selection biases, \citet{sargent_2010} applied a careful
survival analysis to the \textit{Spitzer} (24 and 70$\,\mu$m) and VLA
catalogues.  Finally, we compare our results with those of
\citet{ivison_2010b}.  In this paper, the authors used
\textit{Herschel} observations of the GOODS-N field to constrain the
redshift evolution of $q_{\rm FIR}$ on a relative small sample of
$L_{\rm IR}$-matched galaxies.

Results from the literature are in broad agreement with our
conclusions.  Data from \citet{bourne_2011} exhibit a consistent $q_{\rm
  FIR}\,$--$\,z$ correlation, though with a slight overall offset of
their $q_{\rm FIR}$ values ($\Delta
q_{\rm FIR}$$\,\thicksim\,$$0.15$).
\citet{bourne_2011} also noticed that their $q_{\rm FIR}$ values were systematically higher than that expected from local observations.
They attributed this offset to difference in the assumptions made to infer $L_{\rm IR}$ from \textit{Spitzer} observations.
Most likely, the overall offset observed here between their and our measurements has the same origin.

Results from \citet{ivison_2010b} also exhibit a clear decrease in $q_{\rm FIR}$
with redshift.
However, here as well there exists some disagreement with respect to our measurements: the decrease in $q_{\rm FIR}$ starts at $z>0.7$, and at $0.7$$\,<\,$$z$$\,<\,$$1.4$, their measurements of $q_{\rm FIR}$ are significantly higher.
These discrepancies could not be explained by difference in the assumptions made to derive $L_{\rm FIR}$, because both studies used FIR observations from \textit{Herschel}.
Instead, these discrepancies might be explained by cosmic variance and/or by some differences in our sample selection.
For example, at $z\thicksim0.5$, the measurement of \citet{ivison_2010b} relied on only 16 galaxies and they argued that it might need to be discounted.
In addition, the sample of \citet{ivison_2010b} is $L_{\rm IR}$-selected and not SFR$-M_{\ast}$-selected, and still includes X-ray sources.

The results of \citet{sargent_2010} are
perfectly in line with our findings. They found a clear decrease in $q_{\rm FIR}$ with redshift, and the amplitude and normalisation of their $q_{\rm FIR}\,$--$\,z$ correlation is consistent with our measurements.

From all these comparisons, we conclude that results from the literature are also broadly consistent with a decrease in $q_{\rm FIR}$ with redshift.
Note, however, that these studies have mostly chosen to favour a
non-evolving scenario for the FRC.  This moderate redshift evolution
of $q_{\rm FIR}$, which remains within its local 1$\sigma$ dispersion
even at $z$$\,\thicksim\,$$2$, was not deemed sufficiently significant
in most of these studies, which relied on relatively small samples
with sparse FIR and radio spectral coverage. Our sample is
sufficiently large and well controlled (i.e.\ complete for
star-forming galaxies with $M_{\ast}$$\,>\,$$10^{10}\,$M$_{\odot}$)
with excellent FIR and radio spectral coverage to conclude that
$q_{\rm FIR}$ evolves with redshift as $q_{\rm
  FIR}$$\,\propto\,$$(1+z)^{-0.12\pm0.04}$. Possible physical
explanations and implications of this redshift evolution are discussed
in Sect.~\ref{sec:discussion}.\\

\begin{figure*}
\center
\includegraphics[width=14.9cm]{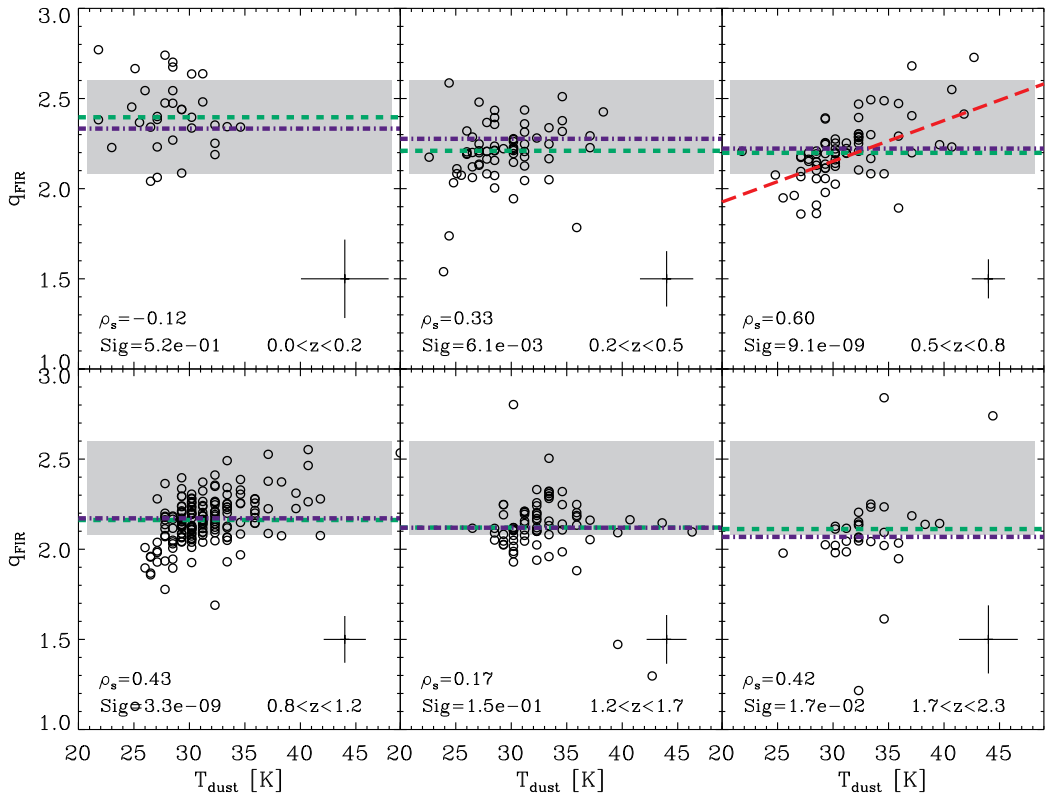}
\caption{ \label{fig:q mstar tdust} Mean $L_{\rm FIR}\,$-- to
  --$\,L_{\rm 1.4\,GHz}$ ratio ($q_{\rm FIR}$) of galaxies as a
  function of dust temperature, as derived from our stacking analysis.
  In each panel, we give the Spearman rank correlation ($\rho_{\rm
    s}$), the null hypothesis probability (Sig.), and show the median
  uncertainties on our $q_{\rm FIR}$ and $T_{\rm dust}$ estimates.  In the 
  redshift bin with a statistically significant (Sig.$\,<\,$$5\%$ and $\chi^{2}_{\rm red}[{\rm linear}]$$\,<\,$$\chi^{2}_{\rm red}[{\rm cst}]$)
  correlation, we plot (red long-dashed lines) a linear fit, $q_{\rm FIR}= (0.023\pm0.008)\times T_{\rm dust}+(1.47\pm0.32)$.
  The rest of the lines and shaded region are the same as in Fig.~\ref{fig:q dssfr}.}
\end{figure*}

In Fig.~\ref{fig:q mstar tdust}, we consider the reality of a
correlation between $q_{\rm FIR}$ and $T_{\rm dust}$.  In all our
redshift bins, there is a weak, positive $q_{\rm FIR}\,$--$\,T_{\rm
  dust}$ correlation; in four bins we can reject the null hypothesis
of uncorrelated data with high significance (Sig.$\,<\,$5\%). 
However, using our Monte-Carlo approach to fit this relation, we find that only at $0.5$$\,<\,$$z$$\,<\,$$0.8$ the linear model has better reduced $\chi^{2}$ values than the constant model and has a slope different than zero at the $\thicksim\,$$3$$\sigma$ level (i.e.\ $\Delta[q_{\rm FIR}]$/$\Delta[T_{\rm dust}]=0.023\pm0.008$).
For the other redshift bins, the linear and constant models are statistically undistinguishable and the slopes of the linear model are consistent, within 1$\sigma$, with zero.
Thus, we conclude that the existence of a $q_{\rm FIR}\,$--$\,T_{\rm dust}$ correlation is statistically meaningful in only one of our redshift bin.
Such $q_{\rm FIR}\,$--$\,T_{\rm dust}$ correlation could be
related to the weak, positive $q_{\rm FIR}\,$--$\,\Delta$log$({\rm
  SSFR})_{\rm MS}$ correlation observed in the same redshift bin,
because $T_{\rm dust}$ is known to be positively correlated with
$\Delta$log$({\rm SSFR})_{\rm MS}$ \citep{magnelli_2014}.  
Note, however, that our estimates of $T_{\rm dust}$ might be affected/contaminated by AGN emission, complicating the interpretation on the existence of a $q_{\rm FIR}\,$--$\,T_{\rm dust}$ correlation.
  
\section{Discussion\label{sec:discussion}}

\subsection{Evolution of the radio spectra\label{subsec:index discussion}}

Our results indicate that the radio spectral index, $\alpha^{\rm
  1.4\,GHz}_{\rm 610\,MHz}$, does not significantly evolve with
redshift, does not correlate significantly with $\Delta$log$({\rm
  SSFR})_{\rm MS}$ and is consistent everywhere with its canonical
value of $0.8$ (see Figs.~\ref{fig:index} and \ref{fig:index z}).  Those results are valid for all star-forming galaxies
with $M_{\ast}$$\,>\,$$10^{10}\,$M$_{\odot}$, $\Delta$log$({\rm
  SSFR})_{\rm MS}$$\,>\,$$-0.3$ and $0$$\,<\,$$z$$\,<\,$$2.3$.

\subsubsection{AGN contamination ?}
Some may find the absence of significant redshift evolution in
$\alpha$ surprising, perhaps anticipating the rapid increase in the
AGN population at $z$$\,>\,$$1$
\citep[e.g.][]{hasinger_2005,wall_2005} and their influence on the observed radio spectral index of their host galaxies.
At low redshift and therefore low rest-frame frequencies, AGNs can
exhibit flat radio spectra \citep[$\alpha$$\,<\,$$0.5$;
e.g.][]{murphy_2013} while at higher redshift and thus higher
rest-frame frequencies they can exhibit steep radio spectra
\citep[$\alpha$$\,>\,$$1.0$; e.g.][]{huynh_2007}. With the increasing
number of AGNs at $z$$\,>\,$$1$, one might then expect variations in
the radio spectral index inferred from our statistical sample.  However, since our goal is
to explore the properties of star-forming galaxies, and anticipating
the possible impact of AGNs, we excluded the brightest X-ray AGNs from
our sample and used a \textit{median} radio stacking method.  This
should have minimised any AGN contamination. 
Interestingly, if we stack our bright X-ray AGNs, we find radio spectra indices in the range [0.7-1.0]\footnote{Precisely, we found $\alpha^{\rm 1.4\,GHz}_{\rm 610\,MHz}$$=1.0\pm0.1,\ 0.7\pm0.1,\, 0.8\pm0.1$ and $0.7\pm0.1$ for X-ray AGNs at $0.5$$\,<\,$$z$$\,<\,$$0.8$, $0.8$$\,<\,$$z$$\,<\,$$1.2$, $1.2$$\,<\,$$z$$\,<\,$$1.7$ and $1.7$$\,<\,$$z$$\,<\,$$2.3$, respectively. Unfortunately, the numbers of X-ray AGNs at $z<0.5$ is not large enough to obtain a meaningful estimate of $\alpha^{\rm 1.4\,GHz}_{\rm 610\,MHz}$ using our sample.}.
This suggests that the radio spectral index of the bulk of the X-rays AGN population is dominated by emission from their host galaxies.
Note, however, that the AGN contribution to the radio emission of a galaxy could also have a radio spectral index of $\thicksim\,$$0.75$.
Indeed, while the radio emission of the compact core of AGNs is supposed to have a flat spectrum at low frequencies and a steep spectrum at high frequencies, their extended radio emission associated with lobes is supposed to have a power-law spectrum with $\alpha$$\,\thicksim\,$$0.75$ \citep{jackson_1999}.

\subsubsection{Nature of the radio spectra\label{subsubsect:nature radio}}
Our VLA and GMRT observations probe different rest-frame radio
frequencies at different redshifts.  Consequently, even if all
star-forming galaxies have the same intrinsic radio spectra, one could
still expect to observe redshift evolution of $\alpha^{\rm
  1.4\,GHz}_{\rm 610\,MHz}$ in the presence of a curved spectrum.  Such
redshift evolution of $\alpha^{\rm 1.4\,GHz}_{\rm 610\,MHz}$ is not
observed in our sample, suggesting that in the range of rest-frame
frequencies probed here, $610\,{\rm MHz}$$\,<\,$$\nu_{\rm
  rest}$$\,<\,$$4.2\,{\rm GHz}$, the radio spectra are well described
by a power-law function with $S_{\nu}$$\,\propto\,$$\nu^{-0.8}$.  The radio spectra of high-redshift star-forming galaxies
seem to be dominated by non-thermal optically thin synchrotron emission with the same properties as that observed in the local Universe;  thermal free-free emission with
relatively flat radio spectra (i.e.\ $\alpha$$\,\thicksim\,$$0.1$)
does not dominate their rest-GHz radio spectra.  Unfortunately, these conclusions might be compromised if the constant $\alpha^{\rm
  1.4\,GHz}_{\rm 610\,MHz}$ value is caused by a `conspiracy',
involving a curved radio spectra and some intrinsic evolution with
redshift.

In the local Universe, starbursts/ULIRGs have significantly flatter radio spectra
($\alpha$$\,\thicksim\,$$0.5$) than normal star-forming galaxies
\citep[$\alpha$$\,\thicksim\,$$0.8$;][]{condon_1991,clemens_2008}.
While our sample might exhibit some flattening of the radio spectral
index when moving from the MS regime ($\Delta$log$({\rm SSFR})_{\rm
  MS}$$\,\thicksim\,$$0$) to the starburst regime ($\Delta$log$({\rm
  SSFR})_{\rm MS}$$\,\thicksim\,$$1$), these $\alpha^{\rm
  1.4\,GHz}_{\rm 610\,MHz}\,$--$\,\Delta$log$({\rm SSFR})_{\rm MS}$
correlations are weak and statistically insignificant (Sig.$\,>\,$$5\%$).  The observation in high-redshift starbursts of
steeper radio spectra than in the local
Universe echoes results for samples of
SMGs, which also have $\alpha$$\,\thicksim\,$$0.8$ \citep{ibar_2010, thomson_2014}.  Indeed, luminous SMGs are
situated well above the MS \citep{magnelli_2012}, are believed to
be strong starbursts and to be the high-redshift counterparts of local ULIRGs.
The fact that high-redshift starbursts have steeper radio spectra suggests that they may not have the same
ISM conditions as their local counterparts, as already hinted by numerous other studies looking at their size
\citep[e.g.][]{tacconi_2006,biggs_2008,farrah_2008} and physical properties
\citep[e.g.][]{ivison_2010c,farrah_2008}.
Alternatively, it could indicate that
the flatter radio spectra of local ULIRGs are due to low-frequency
free-free absorption, as advocated by \citet{condon_1991}, a less
relevant effect at the higher rest-frame frequencies probed at high
redshift, though at $z\thicksim2$ our data are not probing above
$\thicksim$4\,GHz.

\subsubsection{Expectations from theory}
Using a single-zone model of CR injection, cooling and escape,
\citet[][hereafter L10]{lacki_2010a} and \citet[][hereafter
LT10]{lacki_2010b} explore the radio spectra of normal and starburst
galaxies and their evolution with redshift.  At $z$$\,\thicksim\,$$0$,
their model predicts a relatively flat radio spectrum at rest-GHz
frequencies for `compact starbursts' ($\alpha$$\,\thicksim\,$$0.5$) and
a steeper spectrum for `normal star-forming galaxies' ($\alpha$$\,\thicksim\,$$0.8$),
though this difference decreases at higher rest-frame frequencies
($\gg\,$$10\,$GHz) where both asymptote to
$\alpha$$\,\thicksim\,$$1.1$.  The flatter radio spectrum of
starbursts is attributed to more efficient Bremsstrahlung and
ionisation cooling of CR electrons and positrons in their dense ISM.
Thus, the L10 model predicts the observed flatter radio spectra of
local ULIRGs but contradicts the hypothesis of \citet{condon_1991}
about its origin via free-free absorption. 

Our observations do not reveal a significant decrease in the radio
spectral index as we move from normal star-forming galaxies ($\Delta$log$({\rm
  SSFR})_{\rm MS}$$\,\thicksim\,$$0$) to starbursts ($\Delta$log$({\rm
  SSFR})_{\rm MS}$$\,\thicksim\,$$1$).  
  Instead, high-redshift starbursts seem to have roughly the same radio spectral index as normal star-forming galaxies.
  As already mentioned, this result was hinted by the observation of steeper radio spectra and more extended star-forming regions in high-redshift SMGs than in local ULIRGs.
  Therefore, LT10 suggested that high-redshift SMGs have larger CR scale heights ($h$$\,\thicksim\,$$1\,$kpc), consistent with that observed in local normal star-forming galaxies.
This large CR scale height leads to less efficient Bremsstrahlung and ionisation cooling of CRs and thus steeper radio spectra. 
Our results suggest that the bulk of the high-redshift starburst population, and not only SMGs, may have large CR scale heights, labelled `puffy starbursts' in LT10.

LT10 also predict evolution of the radio spectral index of galaxies
with redshift.  Combining intrinsic evolution of the radio spectra
with $k$-correction considerations, they find that $\alpha^{\rm
  1.4\,GHz}_{\rm 610\,MHz}$ of `normal star-forming galaxies' and their `puffy
starbursts' should evolve from $\thicksim\,$0.8 to $\thicksim\,$1.0 between $z$$\,\thicksim\,$$0$
and $z$$\,\thicksim\,$2.  The evolution for `compact starbursts' is
predicted to be weaker, $\alpha^{\rm 1.4\,GHz}_{\rm 610\,MHz}$
increasing from $\thicksim\,$0.5 to $\thicksim\,$0.6 between $z$$\,\thicksim\,$$0$ and
$z$$\,\thicksim\,$$2$.  In both cases the steepening of the radio
spectra is attributed to the increase with redshift in IC losses from
the CMB.  In our analysis we do not find any significant increase in
$\alpha^{\rm 1.4\,GHz}_{\rm 610\,MHz}$ with redshift.  
However, our large measurements errors on $\alpha$ ($\sigma_{\alpha}^{\rm measurements}$$\,\thicksim\,$$0.1-0.2$) and the existence of large intrinsic dispersion as revealed by studies of individually detected high-redshift galaxies \citep[$\sigma_{\alpha}^{\rm intrinsic}$$\thicksim$$0.3$;][]{ibar_2009,ibar_2010} prevents us from ruling out the possibility that high-redshift star-forming galaxies follow the expectations of LT10 for `normal star-forming galaxies' and `puffy starbursts'.
Further insight into any subtle evolution of $\alpha$ with redshift
(as well as with $\Delta$log$({\rm SSFR})_{\rm MS}$) will require
deeper multi-frequency radio observations.  

\subsection{Evolution of the FIR/radio correlation\label{subsec:q discussion}}

Our results indicate that the FRC evolves with redshift as $q_{\rm
  FIR}(z)$$\,=\,$$(2.35\pm0.08)\times(1+z)^{-0.12\pm0.04}$ (see Fig. ~\ref{fig:q z}).
  They also indicate that the FRC does not significantly evolve with $\Delta$log$({\rm SSFR})_{\rm MS}$, though the presence of a weak positive trend, as observed in one of our redshift bin (i.e.\ $\Delta[q_{\rm FIR}]$/$\Delta$[$\Delta$log$({\rm SSFR})_{\rm MS}]=0.22\pm0.07$ at $0.5$$\,<\,$$z$$\,<\,$$0.8$), cannot be firmly ruled out using our dataset (see Fig.~\ref{fig:q dssfr}).
 These
results are valid for all star-forming galaxies with
$M_{\ast}$$\,>\,$$10^{10}\,$M$_{\odot}$, $\Delta$log$({\rm SSFR})_{\rm
  MS}$$\,>\,$$-0.3$ and $0$$\,<\,$$z$$\,<\,$$2.3$.  

\subsubsection{AGN contamination ?}

The presence of a large population of AGNs in our sample at
$z$$\,>\,$$1$ might be a concern for our $q_{\rm FIR}$ estimates.
Anticipating this problem, we have removed X-ray AGNs from our sample
and used a \textit{median} radio stacking analysis to statistically exclude the relatively small population of
radio-loud AGNs.  However, these
precautions will minimise but not completely eliminate contamination
by AGNs.
To further reduce this potential contamination, we repeat our stacking analysis removing from our sample AGNs selected by the IRAC colour-colour criteria of \citet{lacy_2007}\footnote{\citet{donley_2012} have proposed more restrictive IRAC AGN colour-colour criteria to decrease the contamination by star-forming galaxies. However, because the completeness in term of AGN selection of these more restrictive IRAC criteria might be lower, we decided to use the original \citet{lacy_2007} definition.}.
This selection should exclude of our analysis obscured AGNs missed in X-ray observations \citep{lacy_2007,stern_2005,donley_2012}.
 We found $11\,865$ IRAC AGNs in our sample, of which $4\,347$ have individual mid-infrared detections.
 This corresponds to $\thicksim\,$$4\%$ of our final sample and $\thicksim\,$$12\%$ of our final sample with individual mid-infrared detections.
Excluding IRAC AGNs does not change qualitatively and quantitatively our results: $q_{\rm FIR}$ smoothly decreases by $\thicksim\,$$0.3$ across $0$$\,<\,$$z$$\,<\,$$2.3$ and there is no significant $q_{\rm }\,$--$\,\Delta$log$({\rm SSFR})_{\rm MS}$ correlation.
We conclude that the decrease in $q_{\rm FIR}$ is most likely not driven by AGN contamination.
Note that this absence of significant contamination from AGNs echoes results from \citet[][see also Bonzini et al., in prep.]{sargent_2010}.
Indeed, using individually detected FIR and radio sources, \citet{sargent_2010} found that AGNs (X-ray selected and optically selected) and star-forming galaxies follow the same FRC (in terms of normalisation and dispersion) out to at least $z$$\,\thicksim\,$$1.4$.

\subsubsection{Expectations from theory}

Theory predicts that the FRC is driven by star-formation activity in
galaxies.  The UV emission of young, massive ($\gtrsim8\,$M$_{\odot}$)
stars is absorbed by dust and re-emitted in the FIR.  This creates a
linear correlation between SFR and $L_{\rm IR}$, if galaxies are
optically thick at UV wavelengths.  After few Myrs, young massive
stars explode into SNe, accelerating CRs into the general magnetic
field of galaxies and resulting in diffuse synchrotron emission.
Thus, when averaged over the star-formation episode (neglecting the
short lag between the UV emission and explosion of the first massive young
stars), we expect a clear link between SFR, FIR and radio synchrotron
emission from star-forming galaxies.  This constitutes the essence of
the calorimeter theory first proposed in \citet{volk_1989}.  In this
theory, galaxies are both UV and electron calorimeters, i.e.\
\textit{all} UV radiation from young stars is re-emitted by dust in
the FIR and \textit{all} CR electrons are converted within the
galaxies into an observable form, mainly via synchrotron CR cooling.

The calorimeter theory has however been questioned, especially in
light of the remarkably tight FRC over three orders of magnitude in
luminosity.  L10 proposed that -- on top of calorimetry -- a number of additional
physical processes conspire to yield such a tight FRC, covering dwarf
galaxies and ULIRGs.  At low SFR surface density ($\Sigma_{\rm SFR}$\footnote{or
  equivalently low $\Sigma_{\rm gas}$, according to the
  Schmidt-Kennicutt-relation, $\Sigma_{\rm
    SFR}$$\,\propto\,$$\Sigma_{\rm gas}^N$ with $N=1\,$--$\,2$}), the
UV calorimeter assumption fails: galaxies are not optically thick and
some UV photons escape without being re-emitted in the FIR \citep[e.g.][]{buat_2005}.  At the
same time, the CR electrons calorimeter assumption also fails.  CR
electrons escape without radiating their energy in the radio, placing
the galaxies back onto the FRC.  At high $\Sigma_{\rm SFR}$, CR
cooling via Bremsstrahlung, ionisation and IC processes become more
important because of higher gas densities.  CR cooling via synchrotron
competes with these processes, decreasing the radio synchrotron
emission of compact starbursts.  However, because of the higher gas
densities, compact starbursts become CR proton calorimeters.  CR
protons convert their energy via inelastic scattering into gamma rays,
neutrinos and secondary protons and electrons.  These secondary
protons and electrons undergo synchrotron cooling, placing `compact
starbursts' back onto the FRC.  From this, L10 conclude that
calorimetry combines with several conspiracies operating in different
density regimes to produce a relatively constant (variation
$<\,$$0.3$) FRC across the range $0.001\,{\rm
  M_{\odot}\,kpc^{-2}\,yr^{-1}}$$\,<\,$$\Sigma_{\rm
  SFR}$$\,<\,$$1000\,{\rm M_{\odot}\,kpc^{-2}\,yr^{-1}}$.
 
Other theories have been proposed to explain the FRC.  For example
\citet[][hereafter S13; see also Niklas \& Beck
\citeyear{niklas_1997}]{schleicher_2013} explained the FRC by relating
star formation and magnetic field strength via turbulent magnetic
field amplification, the so-called small-scale dynamo effect.  This
model explains the FRC of galaxies both globally and on kiloparsec
scales.  Contrary to the findings of L10, this model predicts that the
FRC should evolve with $\Sigma_{\rm SFR}$ as $q_{\rm
  FIR}$$\,=\,$$-0.3\times{\rm log}(\Sigma_{\rm SFR})+C_1$. \\
 
$\bullet\ $\textit{ Is the FRC expected to evolve with $\Delta$log$({\rm SSFR})_{\rm MS}\,$?\\}
We can relate
$\Delta$log$({\rm SSFR})_{\rm MS}$ with $\Sigma_{\rm SFR}$ using
estimates presented in \citet{wuyts_2011b} and based on almost the same
galaxy sample used here (see Section~\ref{subsec:multi-sample}).
\citet{wuyts_2011b} found that $\Sigma_{\rm SFR}$ increases linearly
with $\Delta$log$({\rm SSFR})_{\rm MS}$.  At $z$$\,\thicksim\,$$0.1$,
MS galaxies ($\Delta$log$({\rm SSFR})_{\rm MS}$$\,\thicksim\,$$0$)
have $\Sigma_{\rm SFR}$$\,\thicksim\,$$0.02\,{\rm
  M_{\odot}\,kpc^{-2}\,yr^{-1}}$ while starbursts ($\Delta$log$({\rm
  SSFR})_{\rm MS}$$\,\thicksim\,$$1$) have $\Sigma_{\rm
  SFR}$$\,\thicksim\,$$0.2\,{\rm M_{\odot}\,kpc^{-2}\,yr^{-1}}$.  At $z$$\,\thicksim\,$$2.0$, MS galaxies have $\Sigma_{\rm
  SFR}$$\,\thicksim\,$$1\,{\rm M_{\odot}\,kpc^{-2}\,yr^{-1}}$ while
starbursts have $\Sigma_{\rm SFR}$$\,\thicksim\,$$10\,{\rm
  M_{\odot}\,kpc^{-2}\,yr^{-1}}$ \citep[for more details, see fig.~4
of][]{wuyts_2011b}.  Thus, in a given redshift bin, our analysis
probes at least one order of magnitude in $\Sigma_{\rm SFR}$.  Across
this range, the model of L10 expects no significant
evolution\footnote{This is true even if high-redshift galaxies follow
  the normal-to-puffy starburst track of LT10. Indeed, at
  $z$$\,\thicksim\,$$2$, our galaxies have $\Sigma_{\rm
    SFR}$$\,>\,$$1\,{\rm M_{\odot}\,kpc^{-2}\,yr^{-1}}$, a
  $\Sigma_{\rm SFR}$ regime where LT10 expect constant $q_{\rm FIR}$.}
of $q_{\rm FIR}$, while the model of S13 expects $q_{\rm FIR}$ to
decrease by $0.3$. Here, we find that, if it exists, a $q_{\rm FIR}\,$--$\,\Delta$log$({\rm SSFR})_{\rm MS}$
correlation is necessary weak and rather positive ($\Delta[q_{\rm FIR}]$/$\Delta$[$\Delta$log$({\rm SSFR})_{\rm MS}]=0.22\pm0.07$).
This result disfavours the model of S13.\\

$\bullet\ $\textit{ Is the FRC expected to evolve with redshift$\,$?\\} 
LT10 and S13 also study the evolution of the FRC with redshift.  At
high redshift the main concern is that other CR cooling processes
might start to dominate over synchrotron.  In particular, IC cooling
from the CMB might become dominant at high redshift ($U_{\rm
  CMB}$$\,\propto\,$$(1+z)^{4}$), leading to a significant increase in
$q_{\rm FIR}$ \citep[see also][]{murphy_2009b}.  However, both studies
concluded that the FRC should hold with no dramatic break-down (i.e.\
$q_{\rm FIR}(z)-q_{\rm FIR}(${\footnotesize
  $z=0$}$)$$\,\gtrsim\,$$0.5$) out to relatively high redshift: up to
$z$$\,\thicksim\,$$2\ (8)$ for galaxies with
$\Sigma$$\,\thicksim\,$$1\,{\rm M_{\odot}\,kpc^{-2}\,yr^{-1}}$ and up
to $z$$\,\thicksim\,$$3\ (15)$ for galaxies with
$\Sigma$$\,\thicksim\,$$10\,{\rm M_{\odot}\,kpc^{-2}\,yr^{-1}}$ in S13
(LT10).  Because at $z$$\,\thicksim\,$$2$ our galaxies have
$\Sigma_{\rm SFR}$$\,>\,$$1\,{\rm M_{\odot}\,kpc^{-2}\,yr^{-1}}$, none
of these models expect a dramatic increase in $q_{\rm FIR}$ for our
sample.  These expectations are in line with our results.

Apart from this potential break-down of the FRC due to IC cooling from
the CMB, LT10 also study the possibility of more subtle variations of
the FRC with redshift.  Because SMGs exhibited low $q_{\rm FIR}$
values \citep[$q_{\rm FIR}$$\,\thicksim\,$$2.0$;][]{murphy_2009} and steep radio spectra \citep[$\alpha$$\,\thicksim\,$$0.8$;][]{ibar_2010}, LT10
postulate that high-redshift SMGs have larger CR scale heights
than their local counterparts ($h$$\,\thicksim\,$$1\,$kpc instead of
$h$$\,\thicksim\,$$0.1\,$kpc).  Then, assuming that the magnetic field
strength varies with $\Sigma_{\rm gas}$ and not $\rho_{\rm gas}$, they
find that such large CR scale heights decrease the CR losses via
Bremsstrahlung and ionisation processes, increasing the synchrotron
emission and decreasing $q_{\rm FIR}$ by $\thicksim\,$$0.3$.
LT10 argue that such `puffy starbursts' are also supported by
kinematic observations of SMGs, citing \citet{tacconi_2006}.  
From this, LT10 predict two different evolution of $q_{\rm FIR}$ across the range $0.001\,{\rm M_{\odot}\,kpc^{-2}\,yr^{-1}}$$\,<\,$$\Sigma_{\rm SFR}$$\,<\,$$1000\,{\rm M_{\odot}\,kpc^{-2}\,yr^{-1}}$.
In their normal-to-compact starburst track, $q_{\rm FIR}$ remains constant as a function of $\Sigma_{\rm SFR}$, while in their normal-to-puffy starburst track $q_{\rm FIR}$ significantly evolves with $\Sigma_{\rm SFR}$.
In this latter, `normal star-forming galaxies' and `puffy starbursts' have the same CR scale height and $q_{\rm FIR}$ smoothly decreases by $0.3$ from $\Sigma_{\rm SFR}$$\,=\,$$0.001\,{\rm M_{\odot}\,kpc^{-2}\,yr^{-1}}$ to $\Sigma_{\rm SFR}$$\,=\,$$1\,{\rm M_{\odot}\,kpc^{-2}\,yr^{-1}}$ and then $q_{\rm FIR}$$\,\thicksim\,$$2$ at $\Sigma_{\rm  SFR}$$\,\gtrsim\,$$1\,{\rm M_{\odot}\,kpc^{-2}\,yr^{-1}}$.
Therefore, while LT10 do not formally predict a particular redshift evolution of $q_{\rm FIR}$, they expect that if the CR scale height of star-forming galaxies evolve with redshifts, $q_{\rm FIR}$ should evolve accordingly.
At $z$$\,\thicksim\,$$2$, all our galaxies have $\Sigma_{\rm SFR}$$\,>\,$$1\,{\rm M_{\odot}\,kpc^{-2}\,yr^{-1}}$ and their $q_{\rm FIR}$ have decreased by $0.3$.  
These observations suggest that most high-redshift star-forming galaxies may have CR scale heights of $\thicksim\,$$1\,$kpc, not only SMGs.
This is in qualitative agreement with the finding that many high-redshift MS star-forming galaxies are clumpy disks with larger disk velocity dispersions (and hence scale heights) than local spirals \citep[e.g.][]{forster_2009,newman_2013}.

S13 also predict some subtle variations of $q_{\rm FIR}$ with redshift
following,
\begin{equation}
 q_{\rm FIR} = -0.3\,{\rm log}(\Sigma_{\rm SFR})-\frac{4\,\beta}{6}\,{\rm log}(1+z)+C_2,
\end{equation}
where $\beta$ parametrises the evolution of the typical ISM density of
galaxies with redshift, $\rho$$\,=\,$$\rho_{0}\,(1+z)^{\,\beta}$.  For
MS galaxies, $\Sigma_{\rm SFR}$ increases by $1.7\,$dex from
$z$$\,\thicksim\,$$0$ to $z$$\,\thicksim\,$$2$.  Assuming that their
CR scale heights stay the same across this redshift range
($h$$\,\thicksim\,$$1\,$kpc) and that $\Sigma_{\rm
  SFR}$$\,=\,$$\Sigma_{\rm gas}^{N}$$\,\thicksim\,$$(h\,\rho)^N$ with
$N$$\,\thicksim\,$$1\,$--$\,2$ from the Schmidt-Kennicutt relation,
one would infer that $\rho_{z\thicksim2}^{\rm
  MS}$$\,=\,$$\rho_{z=0}^{MS}\,(1+z)^{\,\beta_{\rm MS}}$ with
$\beta_{\rm MS}$$\,\thicksim\,$$1.8\,$--$\,3.6$. Using the model of
S13, we therefore predict that $q_{\rm FIR}$ for MS galaxies should
decrease by $-1.0\,$($-1.6$) between $z$$\,\thicksim\,$$0$ and
$z$$\,\thicksim\,$$2$ for $N$$\,=\,$$2$~$(1)$.  Such large evolution
of $q_{\rm FIR}$ is not supported by our observations.  Similarly, one
can predict the evolution of $q_{\rm FIR}$ for starbursts galaxies
($\Delta$log$({\rm SSFR})_{\rm MS}\gtrsim0.75$).  Their $\Sigma_{\rm
  SFR}$ also increase by $1.7\,$dex from $z$$\,\thicksim\,$$0$ to
$z$$\,\thicksim\,$$2$.  Assuming that their CR scale heights evolve
from $h$$\,\thicksim\,$$0.1\,$kpc to $h$$\,\thicksim\,$$1\,$kpc from
$z$$\,\thicksim\,$$0$ to $z$$\,\thicksim\,$$2$, we predict -- using
the model of S13 -- a decrease in $q_{\rm FIR}$ by
$-0.4\,$($-1.0$) for $N$$\,=\,$$2$~$(1)$ across this redshift
range\footnote{Assuming that the CR scale height of starbursts is
  $0.1\,$kpc across $0$$\,<\,$$z$$\,<\,$$2.3$, we infer similar
  evolution of $q_{\rm FIR}$ than for MS galaxies, i.e.\
  $-1.0$$\,<\,$$\Delta q_{\rm FIR}$$\,<\,$$-1.6$.}.  Again, such
large evolution of $q_{\rm FIR}$ is not supported by our observations.\\

$\bullet\ $\textit{ Is the FRC expected to evolve with redshift in the context of the MS of star formation$\,$?\\}
 There is strong observational
evidence that the physical conditions reigning in the star-forming
regions of high-redshift MS galaxies are similar to those of normal
local star-forming galaxies, though high-redshift MS galaxies have
larger $\Sigma_{\rm SFR}$ \citep{wuyts_2011b, elbaz_2011, nordon_2012,magnelli_2012b,magnelli_2014}.
Thus, MS galaxies should follow
the normal-to-puffy starburst track of LT10 across $0$$\,<\,$$z$$\,<\,$$2.3$, with its constant CR
scale height of $\thicksim\,$$1\,$kpc.  This would translate into a
smooth decrease in $q_{\rm FIR}$ by $0.3$ at
$z$$\,\thicksim\,$$2$ as their $\Sigma_{\rm SFR}$ varies from
$\Sigma_{\rm SFR}$$\,\thicksim\,$$0.02\,{\rm
  M_{\odot}\,kpc^{-2}\,yr^{-1}}$ to $\Sigma_{\rm
  SFR}$$\,\thicksim\,$$1\,{\rm M_{\odot}\,kpc^{-2}\,yr^{-1}}$ across
this redshift range.
Such predictions are supported by our observations.

In contrast, the decrease in $q_{\rm FIR}$ with redshift for far-above MS galaxies (those with $\Delta$log$({\rm SSFR})_{\rm MS}$$\,\gtrsim\,$$0.75$, see Sect.~\ref{subsec:q}) is more surprising.
Indeed, at high redshift, far-above MS galaxies are thought to be starbursts with similar FIR properties to local ULIRG \citep[e.g.][]{elbaz_2011,magnelli_2012b,magnelli_2014}.  
 Consequently, one could
expect these galaxies to follow the `compact starburst' predictions of LT10
across $0$$\,<\,$$z$$\,<\,$$2.3$, characterised by a constant value of
$q_{\rm FIR}$ over a large range of $\Sigma_{\rm SFR}$.  Instead, their
$q_{\rm FIR}$ seems to decrease by $\thicksim\,$$0.3$ at
$z$$\,\thicksim\,$$2$, suggesting that most high-redshift starbursts (and not only SMGs) have larger CR scale heights
(`puffy starbursts') than their local counterparts (`compact starbursts').
Consequently, while high-redshift starbursts seems to share some physical properties with local ULIRGs (FIR SEDs; CO-to-H$_{2}$ conversion factors), other properties seem to be significantly different (size of their star-forming regions and CR scale heights).
Note, however, that there is a larger dispersion of $q_{\rm FIR}$ at high $\,\Delta$log$({\rm SSFR})_{\rm MS}$ values and that the $q_{\rm FIR}\,$--$\,z$ correlation is statistically less significant  for far-above MS galaxies than for MS galaxies ($q_{\rm FIR}[z,\,$far-above~MS]$\,\propto\,$$(1+z)^{-0.13\pm0.09}$ and $q_{\rm FIR}[z,{\rm MS}]$$\,\propto\,$$(1+z)^{-0.16\pm0.05}$).
This supports a more complex scenario in which a non-negligible number of high-redshift
starbursts are associated with `compact starbursts' rather than with `puffy starbursts'.

\subsubsection{Implications of the evolution with redshifts of the FRC}

Our results have important implications for previous and future studies relying on the FRC.  For
example, SFRs of high-redshift galaxies determined via radio
observations will be lower than expected from the local FRC.  However,
while the evolution of $q_{\rm FIR}$ with redshift is statistically
significant, it is modest and still lies within its local 1$\sigma$
dispersion at $z$$\,\thicksim\,$$2$.  Thus, corrections of SFRs
motivated by our findings should still be within the uncertainties
inferred using the local 1$\sigma$ dispersion of the FRC.

Our results have also important implications for our understanding of the contribution of star-forming galaxies to the radio extragalactic background.
In particular, an extragalactic 3.3-GHz radio background of $\nu
I_{\nu}$$\,\approx\,$$5.9\times10^{-4}\,$nW$\,$m$^{-2}\,$sr$^{-1}$ has
been detected with ARCADE2 \citep{fixsen_2011, seiffert_2011}.  The
sources of this cosmic radio background (CRB) have been extensively studied in 
the recent years \citep[e.g.][]{gervasi_2008,singal_2010,vernstrom_2011,ysard_2012,condon_2012,vernstrom_2014}.  All these studies report
that only a small fraction of this background ($\thicksim\,$10--26\%)  can be accounted
for by objects resolved in the deepest existing radio surveys.
Thus, they conclude that if the CRB at $3.3$-GHz is at the level reported by ARCADE2, it must originate from very faint radio sources still to be observed.
In particular, \citet{singal_2010} speculate that these sources could have radio flux densities $<\,$$10\,\mu$Jy and 
be ordinary star-forming galaxies at $z$$\,>\,$$1$
associated with an FRC that has evolved towards radio-loud
(numerically lower) values.  Such evolution of the FRC is qualitatively
supported by our observations.  However, quantitatively, the evolution
of $q_{\rm FIR}$ required by \citet{singal_2010} is much larger than
that observed here, i.e.\ $\Delta q_{\rm FIR}$$\,<\,$$-0.7$ instead of
the observed $\Delta q_{\rm FIR}$$\,\thicksim\,$$-0.3$ at
$z$$\,\thicksim\,$$2$.  From the total infrared luminosity density across
$0$$\,<\,$$z$$\,<\,$$4$ inferred by \citet[][see also Magnelli et al.
\citeyear{magnelli_2013}]{gruppioni_2013} using \textit{Herschel}
observations\footnote{Note that the cosmic infrared background derived
  here from the total infrared luminosity density across
  $0$$\,<\,$$z$$\,<\,$$4$ is of $25.5\,$nW$\,$m$^{-2}\,$sr$^{-1}$, in
  agreement with estimates of \citet{dole_2006} and
  \citet{bethermin_2012}.}, we can estimate the contribution of star-forming galaxies to the CRB at 3.3-GHz.
Assuming $S_{\nu}$$\,\propto\,$$\nu^{-0.8}$ and $q_{\rm
  FIR}$$\,=\,$$2.34$ across $0$$\,<\,$$z$$\,<\,$$4$, we find $\nu
I_{\nu}^{\rm
  FRC}$$\,\approx\,$$3.1\times10^{-5}\,$nW$\,$m$^{-2}\,$sr$^{-1}$,
i.e.\ $5.2\%$ of the total CRB at $3.3\,$GHz.  Instead, using $q_{\rm
  FIR}(z)$$\,=\,$$2.35\times(1+z)^{-0.12}$ at
$0$$\,<\,$$z$$\,<\,$$2.3$ and $q_{\rm
  FIR}(${\footnotesize$z>2$}$)$$\,=\,$$q_{\rm FIR}(${\footnotesize
  $z=2$}$)$, we find $\nu I_{\nu}^{\rm
  FRC}$$\,\approx\,$$4.5\times10^{-5}\,$nW$\,$m$^{-2}\,$sr$^{-1}$,
i.e.\ $7.6\%$ of the total CRB at $3.3\,$GHz.  Finally, using $q_{\rm
  FIR}(z)$$\,=\,$$2.35\times(1+z)^{-0.12}$ at $0$$\,<\,$$z$$\,<\,$$4$,
we find $\nu I_{\nu}^{\rm
  FRC}$$\,\approx\,$$4.6\times10^{-5}\,$nW$\,$m$^{-2}\,$sr$^{-1}$,
i.e.\ $7.7\%$ of the total CRB at $3.3\,$GHz.  Therefore, while the
evolution of the FRC with redshift increases the `resolved' fraction
of the CRB as measured by ARCADE2, it cannot entirely explain its
origin.  We conclude that if the CRB is at the level reported here,
the contribution of star-forming galaxies that obey the FRC is at
most $\thicksim\,$$10\%$ (taking into account a dispersion of the FRC
of $0.3$).
However, note that a proper treatment of outliers of
the FRC should be performed in order to estimate the entire contribution of
infrared sources to the CRB.

We also estimate the contribution of X-ray AGNs to the CRB  at $3.3\,$GHz.
In each of our redshift bin, we measure the mean $1.4$-GHz flux density of X-ray AGNs using our stacking analysis. 
Then, to obtain the $1.4$-GHz radio background of X-ray AGNs, we multiply this mean radio flux by the density of X-ray AGNs detected in our GOODS-N/S samples (i.e.\ fields with the deepest X-ray datasets). 
Finally, we convert this $1.4$-GHz radio background into a $3.3$-GHz radio background assuming a radio spectral index of X-ray AGNs in the range [$0.5\,$--$\,1.0$]. 
We found that X-ray AGNs have $\nu I_{\nu}^{\rm X-ray\ AGNs}$$\,\approx\,$$11.0\ (7.2) \times10^{-6}\,$nW$\,$m$^{-2}\,$sr$^{-1}$, assuming a radio spectral index of 0.5 (1.0).
This corresponds to only $\thicksim\,$$1.9\,\%\ (1.2\,\%)$ of the total CRB measured by ARCADE2 at $3.3$-GHz.

\section{Summary\label{sec:conclusion}}

In this paper we study the evolution of the FRC and radio spectral
index ($\alpha^{\rm 1.4\,GHz}_{\rm 610\,MHz}$) across the
SFR$-M_{\ast}$ plane and up to $z$$\,\thicksim\,$$2$.  We use the
deepest FIR-\textit{Herschel}, 1.4-GHz (VLA) and 610-MHz (GMRT)
observations available for the GOODS-N, GOODS-S, ECDFS and COSMOS
fields.  Infrared luminosities are inferred using the stacked
PACS/SPIRE FIR photometry for each SFR--$M_{\ast}$--$\,z$ bin.  Radio
luminosities and radio spectral indices are derived using their stacked
1.4-GHz and 610-MHz flux densities.  Using this methodology, we are
able to overcome most of the biases affecting previous studies on the
FRC.  Selection biases are overcome by the use of a complete
stellar-mass-selected ($M_{\ast}\gtrsim10^{10}\,$M$_{\odot}$) sample of
star-forming galaxies at $0$$\,<\,$$z$$\,<\,$$2.3$.  Observational
biases, introduced by sparse coverage of the FIR and radio spectra,
are overcome by the use of multi-wavelength FIR and radio
observations.  Our results, which are valid for all star-forming
galaxies with $M_{\ast}$$\,>\,$$10^{10}\,$M$_{\odot}$,
$\Delta$log$({\rm SSFR})_{\rm MS}$$\,>\,$$-0.3$ and
$0$$\,<\,$$z$$\,<\,$$2.3$, can be summarised as follows:
\begin{enumerate}
\item The radio spectral index, $\alpha^{\rm 1.4\,GHz}_{\rm
    610\,MHz}$, does not evolve significantly with the distance of a
  galaxy with respect to the MS (i.e.\ $\Delta$log$({\rm SSFR})_{\rm
    MS}$) nor with redshift.  Instead, $\alpha^{\rm 1.4\,GHz}_{\rm
    610\,MHz}$ remains relatively constant across the
  SFR--$M_{\ast}$--$\,z$ parameter space, consistent with a canonical
  value of $0.8$.  This results suggests that the radio spectra of the
  bulk of the high-redshift star-forming galaxy population is
  dominated by non-thermal optically thin synchrotron emission, well described by a
  power-law function, $S_{\nu}$$\,\propto\,$$\nu^{-0.8}$, across the
  range of rest-frequencies probed here, $610\,{\rm
    MHz}$$\,<\,$$\nu_{\rm rest}$$\,<\,$$4.2\,{\rm GHz}$.
\item A relatively constant radio spectral index from normal
  star-forming galaxies ($\Delta$log$({\rm SSFR})_{\rm
    MS}$$\,\gtrsim\,$$0$) to starbursts ($\Delta$log$({\rm SSFR})_{\rm
    MS}$$\,\gtrsim\,$$1$) is surprising in light of local observations
  where the radio spectra of ULIRGs are significantly flatter
  ($\alpha$$\,\thicksim\,$$0.5$) than those of spiral galaxies
  \citep[$\alpha$$\,\thicksim\,$$0.8$;][]{condon_1991,clemens_2008}.
  However, the observation of relatively steep radio spectra in
  high-redshift starbursts is also supported by results from
  \citet{ibar_2009} and \citet{thomson_2014} who found
  $\alpha$$\,\thicksim\,$$0.8$ for samples of individually detected
  high-redshift SMGs.  The combination of these results suggests that
  most high-redshift starbursts have different ISM properties (e.g.\ 
  magnetic field strength, gas densities, $\Sigma_{\rm SFR}$, \dots)
  than their local counterparts.  Alternatively, it could suggest
  that, as advocated by \citet{condon_1991}, the flatter radio
  spectrum of local starbursts is due to free-free absorption, less
  relevant at the higher rest-frequencies probed at high redshift.
\item The FRC does not evolve significantly with $\Delta$log$({\rm SSFR})_{\rm MS}$, though the presence of a weak positive trend, as observed in one of our redshift bin (i.e.\ $\Delta[q_{\rm FIR}]$/$\Delta$[$\Delta$log$({\rm SSFR})_{\rm MS}]=0.22\pm0.07$ at $0.5$$\,<\,$$z$$\,<\,$$0.8$), cannot be firmly ruled out using our dataset.
\item The FRC evolves with redshift as $q_{\rm
    FIR}(z)$$\,=\,$$(2.35\pm0.08)\times(1+z)^{-0.12\pm0.04}$.
     This redshift evolution of the FRC is consistent with previous findings from the
  literature, though most favoured a non-evolving FRC because
  high-redshift measurements were within its local 1$\sigma$
  dispersion, $q_{\rm
    FIR}(${\footnotesize$z\approx0$}$)$$\,=\,$$2.34\pm0.26$
  \citep{yun_2001}.  
\item The fact that the FRC still holds at high redshift, albeit with
  some moderate evolution, suggests that IC cooling of CR electrons
  and protons off photons from the CMB ($U_{\rm
    CMB}$$\,\propto\,$$(1+z)^{4}$) does not yet dominate over
  synchrotron cooling at $z$$\,\thicksim\,$$2$.  The redshift
  evolution of the FRC suggests that the ISM properties (e.g.\ 
  magnetic field strength, gas densities, $\Sigma_{\rm SFR}$, \dots)
  of star-forming galaxies evolve between $z$$\,\thicksim\,$$0$ and
  $z$$\,\thicksim\,$$2$.
\end{enumerate}

A decrease in $q_{\rm FIR}$ with redshift was expected by some of the
most up-to-date theoretical models of the FRC, e.g.\ L10, LT10 and
S13.  Such evolution is expected because $\Sigma_{\rm SFR}$ and
$\rho_{\rm gas}$, which control in part the normalisation of the FRC,
are known to evolve significantly with redshift.  While the model of
S13 seems to over-estimate the decrease in $q_{\rm FIR}$ with
redshift, the bulk of the high-redshift star-forming galaxy population
follows the expectations of LT10 by evolving smoothly from their
normal-to-compact starbursts track at $z$$\,\thicksim\,$$0$ to their
normal-to-puffy starbursts track at $z$$\,\thicksim\,$$2$.  This
suggests that MS galaxies have a constant CR scale height of
$\thicksim\,$$1\,$kpc across $0$$\,<\,$$z$$\,<\,$$2.3$ and that their
$q_{\rm FIR}^{\rm MS}$ decrease by $0.3$ at $z$$\,\thicksim\,$$2$
because their $\Sigma_{\rm SFR}$ changes from $\Sigma_{\rm
  SFR}$$\,\thicksim\,$$0.02\,{\rm M_{\odot}\,kpc^{-2}\,yr^{-1}}$ to
$\Sigma_{\rm SFR}$$\,\thicksim\,$$1\,{\rm
  M_{\odot}\,kpc^{-2}\,yr^{-1}}$ in this redshift range.  A constant
CR scale height for MS galaxies is consistent with the current
interpretation of the MS of star formation.  Indeed, despite their
extreme $\Sigma_{\rm SFR}$, high-redshift MS galaxies have the
physical properties of local normal star-forming galaxies
\citep[e.g.][]{wuyts_2011b, elbaz_2011, magnelli_2012b,
  magnelli_2014}.  In addition, our results suggests that most
starbursts have a CR scale height of $\thicksim\,$$0.1\,$kpc at
$z$$\,\thicksim\,$$0$ and of $\thicksim\,$$1\,$kpc at
$z$$\,\thicksim\,$$2$.  This translates into a smooth decrease in
$q_{\rm FIR}$ by $0.3$ at $z$$\,\thicksim\,$$2$, as starbursts
move from the `compact starburst' to the `puffy starburst' predictions of
LT10. Nevertheless, this vision of an homogeneous high-redshift
starburst population, constituted only of `puffy starbursts', is
compromised by the possibility of a weak $q_{\rm
  FIR}\,$--$\,\Delta$log$({\rm SSFR})_{\rm MS}$ correlation, combined
with a larger dispersion in $q_{\rm FIR}$ at high $\,\Delta$log$({\rm
  SSFR})_{\rm MS}$.  These latter observations support a scenario in
which a non-negligible number of high-redshift starbursts are `compact'.

Our results have important implications for studies relying on the
local FRC.  For example, SFRs of high-redshift galaxies determined via
radio observations will be lower than predicted by the local FRC.
However, because $q_{\rm FIR}$ is still within its local 1$\sigma$
dispersion at $z$$\,\thicksim\,$$2$, uncertainties on previous SFR
estimates should capture this evolution.

It has been postulated in the past that the CRB detected by ARCADE2
at $3.3\,$GHz could be dominated by $z$$\,>\,$$1$ star-forming
galaxies obeying an evolved, radio-loud FRC \citep[e.g.][]{singal_2010}.
However, we find that star-forming galaxies responsible for the CIB
contribute at most $\thicksim\,$$10\%$ of the CRB, even after having
accounted for the evolution of the FRC observed here.

\begin{acknowledgements}
  We thank the anonymous referee for suggestions which greatly enhanced this work.
  PACS has been developed by a consortium of institutes led by MPE
  (Germany) and including UVIE (Austria); KU Leuven, CSL, IMEC
  (Belgium); CEA, LAM (France); MPIA (Germany); INAF-IFSI/OAA/OAP/OAT,
  LENS, SISSA (Italy); IAC (Spain). This development has been
  supported by the funding agencies BMVIT (Austria), ESA-PRODEX
  (Belgium), CEA/CNES (France), DLR (Germany), ASI/INAF (Italy), and
  CICYT/MCYT (Spain).  SPIRE has been developed by a consortium of
  institutes led by Cardiff University (UK) and including University
  of Lethbridge (Canada), NAOC (China), CEA, LAM (France), IFSI,
  University of Padua (Italy), IAC (Spain), Stockholm Observatory
  (Sweden), Imperial College London, RAL, UCL-MSSL, UKATC, University
  of Sussex (UK), Caltech, JPL, NHSC, University of Colorado (USA).
  This development has been supported by national funding agencies:
  CSA (Canada); NAOC (China); CEA, CNES, CNRS (France); ASI (Italy);
  MCINN (Spain); SNSB (Sweden); STFC, UKSA (UK); and NASA (USA).
  Support for BM was provided by the DFG priority programme 1573 ``The physics of the interstellar medium''.
  RJI acknowledges support from the European Research Council in the form
  of Advanced Grant, {\sc cosmicism}.
  EI acknowledges funding from CONICYT/FONDECYT postdoctoral
project N$^\circ$:3130504.
FB and AK acknowledge support by the Collaborative Research Council 956, sub-project A1, funded by the Deutsche Forschungsgemeinschaft (DFG).
\end{acknowledgements}
\bibliographystyle{aa}

\end{document}